# Stochastic earthquake source model:

# the omega-square hypothesis and the directivity effect

**G.Molchan**

*Institute of Earthquake Prediction Theory and Mathematical Geophysics,*

*84/32, Profsoyuznaya Str., Moscow, 117997, Russian Federation.*

*The Abdus Salam International Centre for Theoretical Physics, Trieste, Italy.*

*Abbreviated title:* **Stochastic earthquake source model**

*Name and contact details*: George Molchan ,

Fax: +7 495 333 41 24;

E-mail: molchan@mitp.ru

***Abstract.*** Recently A. Gusev suggested and numerically investigated the doubly stochastic earthquake source model. The model is supposed to demonstrate the following features in the far-field body waves: 1) the omega-square high-frequency (HF) behavior of displacement spectra; 2) lack of the directivity effect in HF radiation; and 3) a stochastic nature of the HF signal component. The model involves two stochastic elements: the local stress drop (SD) on a fault and the rupture time function (RT) with a linear dominant component. The goal of the present study is to investigate the Gusev model theoretically and to find conditions for (1, 2) to be valid and stable relative to receiver site. The models with smooth elements SD, RT are insufficient for these purposes. Therefore SD and RT are treated as realizations of stochastic fields of the fractal type. The local smoothness of such fields is characterized by the fractional (Hurst) exponent H, $0 < H < 1$. This allows us to consider a wide class of stochastic functions without regard to their global spectral properties. We show that the omega-square behavior of the model is achieved approximately if the rupture time function is almost regular (H~1) while the stress drop is rough function of any index H. However, if the rupture front is linear, the local stress drop has to be function of minimal smoothness (H~0). The situation with the directivity effect is more complicated: for different RT models with the same fractal index, the effect may or may not occur. The nature of the phenomenon is purely analytical. The main controlling factor for the directivity is the degree of smoothness of the two dimensional distributions of RT random function. For this reason the directivity effect is unstable. This means that in practice the opposite conclusions relative to the statistical significance of the directivity effect are possible.

## 1. Introduction.

One important problem in the study of the earthquake source is to clarify the mechanism of formation of $\omega^{-2}$ high-frequency (HF) behavior of body wave displacement spectra in the far field (Gusev, 2013). Paucity of knowledge on the earthquake source and the needs of engineering seismology give rise to considerably many source models (see surveys by Gusev (2011, 2013)). Gusev summarized specific properties of HF radiation as follows:

(a)there is a plateau in the source acceleration spectrum ('$\omega$-square behavior') bounded by two specific cutoff frequencies with the upper cutoff in the range between 3 and 30 Hz, but often higher, beyond this range;

(b) lack of the directivity effect in HF radiation;

(c) there are fractal and stochastic elements in the structure of radiated HF body waves.

In addition, it is assumed that two different space-time scales exist such that at the macro-scale one can see a simply connected rupture with the front in the form of a smooth line, like a crack tip that propagates in a locally unilateral way. At the micro-scale, the rupture front can be visualized as a multiply connected fractal line or polyline. It propagates locally, in random directions, and is governed by stochastic regularities, including a fractal time structure. A model of this type was proposed by Gusev (2011, 2013, 2014) It includes a random stress drop field and a random rupture time function.

The goal of the present study is to investigate the Gusev model theoretically and to find conditions under which properties (a, b) are valid and stable relative to the receiver position. We are going to show that smooth characteristics of the model are insufficient to provide the stable $\omega$-square behavior. Considering the non-smooth case, the problem is reduced to the search for a link between the fractional smoothness of the components with the exponent of the displacement spectra decreasing. Results of this kind throw light on the numerical results by Gusev (2014), on the omega-square hypothesis, and on the debatable problem of the rupture directivity effect.

## 2. The Problem

Das and Kostrov (1983) considered the problem of spontaneous shear rupture of a single circular asperity $\Sigma$ on an infinite fault plane. The far-field displacement at $(\mathbf{x}_{rec}, t)$ for P, SV, and SH waves is given by

$$u(t) = A\int_{\Sigma} \tau(\mathbf{x}, t - |\mathbf{x} - \mathbf{x}_{rec}|/c)\sigma(d\mathbf{x}) \approx A\int_{\Sigma} \tau(\mathbf{x}, t - t_0 + (\mathbf{x}, \boldsymbol{\gamma})/c)\sigma(d\mathbf{x}) \tag{1}$$

where $\mathbf{x} = (x', x'')$ is the 2-D vector of coordinates on the fault, $\tau(\mathbf{x}, t)$ is a (local) stress drop on a fault-plane element $d\mathbf{x} = dx' \times dx''$ at time t, $\boldsymbol{\gamma} = (\mathbf{x}_{rec} - \mathbf{x}_{hyp})/|\mathbf{x}_{rec} - \mathbf{x}_{hyp}|$ is the hypocentre ($\mathbf{x}_{hyp} = 0$) to receiver ($\mathbf{x}_{rec}$) direction, $c$ is the wave velocity, $(\mathbf{x}, \boldsymbol{\gamma})$ denotes a scalar product, $t_0 = |\mathbf{x}_{rec} - \mathbf{x}_{hyp}|/c$, $\sigma(d\mathbf{x}) = dx'dx''$ and the *A* factor combines constant coefficients, geometric spreading, and the wave radiation pattern for unit force.

Das and Kostrov (1986) modified this approach to apply it to the case of a small asperity on a finite-size frictionless fault. Boatwright (1988) and Gusev (1989) noted that the finite fault theory of Das and Kostrov (1986) can be further generalized: any part of this fault can be treated as an asperity Applying this approach, the small asperity of Das and Kostrov (1983, 1986) will be treated as an element $d\mathbf{x}$ of fault patch, and the integration in (1) will be expanded to cover the entire earthquake fault patch to be denoted $\Sigma$ again. More discussion of (1) can be found in (Gusev, 2013, 2014).

Formally (1) still stands but with a different factor A, if $\tau(\mathbf{x},t)$ is replaced with a local displacement jump rate $\Delta\dot{u}(\mathbf{x},t)$ (Aki and Richards, 1980). Both formulas reflect the dual treatments of the earthquake source based on the notion of either dislocation or of local stress drop. Viewed mathematically, the asymptotic expression looks identically. However, further simplifications of the source function are not always justified in both cases simultaneously (Gusev, 2013).

Suppose that the stress on the fault is released instantly and completely at the time when the rupture front $t_r(\mathbf{x})$ arrives. Then the following simple model of wave velocity is acceptable (see e.g., Gusev, 2013):

$$u(t) = A\int_{\Sigma} \tau(\mathbf{x})\chi(t - t_a(\mathbf{x}))\sigma(d\mathbf{x}), \qquad (2)$$

where $\tau(\mathbf{x})$ is (local) stress drop on $d\mathbf{x}$ at any moment t, $\chi(\cdot)$ is the Heaviside step function and $t_a(\mathbf{x}) = t_0 - (\mathbf{x},\boldsymbol{\gamma})/c + t_r(\mathbf{x})$ is the arrival time at the receiver site along the ray $\boldsymbol{\gamma}$ of a signal emitted by point $\mathbf{x}$ It is assumed that $\tau(\mathbf{x}) > 0$, thus (2) is a unipolar pulse.

In terms of the Fourier transform, relation (2) can be written as

$$-i\omega\hat{u}(\omega) = A\int_{\Sigma} \tau(\mathbf{x})\exp(i\omega t_a(\mathbf{x}))\sigma(d\mathbf{x}). \qquad (3)$$

We assume that the rupture is mostly unilateral with a nearly straight rupture front, and that

$$t_r(\mathbf{x}) = (\mathbf{x},\boldsymbol{\gamma}_r)/v + \delta t_r(\mathbf{x}), \qquad (4)$$

where $\boldsymbol{\gamma}_r$ and $v \neq c$ are constants that represent the dominant direction and velocity of rupture propagation respectively. Then

$$t_a(\mathbf{x}) = t_0 - (\mathbf{x},\mathbf{d}) + \delta t_r(\mathbf{x}), \qquad (5)$$

where $\mathbf{d}$ is the *directivity vector:*

$$\mathbf{d} = \boldsymbol{\gamma}/c - \boldsymbol{\gamma}_r/v. \qquad (6)$$

The original problem includes two points: the order of the asymptotics of the rapidly oscillating integral (3) as $\omega \to \infty$, and the requirement that $\hat{u}(\omega)$ asymptotics should be independent of the directivity $\mathbf{d}$ .

The theory of rapidly oscillating integrals is well developed for sufficiently smooth integrands and a piecewise smooth boundary $\partial\Sigma$ (see, e.g., Erdelyi (1955), Fedoryuk (1987)). In this case the general theory states the following:

**2.1** Assume that $\mathbf{x}_0$ is a regular inner point of $\Sigma$ , i.e. the gradient $\nabla t_a(\mathbf{x}_0) \neq 0$. Then the contribution $\hat{u}_{\mathbf{x}_0}(\omega)$ of this point (with its vicinity) into the $\hat{u}(\omega)$ asymptotics is negligibly small, i.e., $\hat{u}_{\mathbf{x}_0}(\omega) = O(\omega^{-N})$, where N+1 is the degree of differentiability of $\tau_0(\mathbf{x})$

The contribution due to $\mathbf{x}_0$ is given by (3), if we replace $\tau(\mathbf{x})$ by $\tau_0(\mathbf{x}) = \tau(\mathbf{x})\varphi(\mathbf{x} - \mathbf{x}_0)$ , where $\varphi$ is an infinitely smooth function that is 1 near $\mathbf{0}$ and vanishes outside a small vicinity of $\mathbf{0}$ The contribution of $\mathbf{x}_0$ in the HF asymptotics is independent of $\varphi$ .

To explain the statement, note that $t_a(\mathbf{x})$ is locally linear. For this reason the right-hand side of (3) is, roughly speaking, the Fourier transform of a smooth finite function $\tau_0(\mathbf{x})$. In this case it must decay in a power-law manner with the exponent N, where N+1 is the degree of differentiability of $\tau_0(\mathbf{x})$ (Gel'fand, Shilov, 1964). A rigorous proof of **2.1** can be found, e.g., in Fedoryuk (1987) (Ch.3, Lemma 2.1).

**2.2** Assume that $\mathbf{x}_0 = (x', x'')_0$ is an isolated non-degenerate stationary point of $t_a(\mathbf{x})$:

$$\nabla t_a(\mathbf{x}_0) = 0, \qquad \det[\partial_{x^i}\partial_{x^j} t_a(\mathbf{x}_0)] \doteq K(\mathbf{x}_0) \neq 0 \tag{7}$$

($K(\mathbf{x}_0)$ is the total or Gaussian curvature of $t_a(\mathbf{x})$ at $\mathbf{x}_0$; '$\doteq$' means 'by definition ').

Then the stationary point $\mathbf{x}_0$ yields the following contribution into the asymptotics of $\hat{u}(\omega)$:

$$\left|\hat{u}_{\mathbf{x}_0}(\omega)\right| = \omega^{-2} 2\pi \left|A\tau(\mathbf{x}_0)\right| / \left|K(\mathbf{x}_0)\right|^{1/2}, \tag{8}$$

i.e., the stationary point provides the desired contribution into the HF displacement spectrum, viz. $O(\omega^{-2})$

By (5), the stationary points and directivity vector are connected via $\nabla\delta t(\mathbf{x}_0) = \mathbf{d}$. However, this equation may lack any solution when the azimuth of receiver site is changed by $180^0$. For example, if $\delta t_r(\mathbf{x}) = \varepsilon|\mathbf{x}|^2/2$, then $\mathbf{x}_0 = \mathbf{d}/\varepsilon$. But $\mathbf{x}_0$ is the stationary point, if $\mathbf{d}/\varepsilon \in \Sigma$ Therefore, the role of isolated non-degenerate stationary points in the generation of the $\omega^{-2}$ behavior of $\hat{u}(\omega)$ is unstable, i.e., stationary points can disappear and reappear both either after small perturbations of $\delta t_r(\mathbf{x})$ or after changes of the ray $\boldsymbol{\gamma}$ .

We now explain (8). Under **2.2,** $t_a(\mathbf{x})$ is approximately a quadratic function in a vicinity of $\mathbf{x}_0$. Therefore, taking a suitable coordinates $(u,v)$ with the origin at $\mathbf{x}_0$, $t_a(\mathbf{x})$ can be represented as $au^2 + bv^2$. Since we have a degree of freedom in our choice of $\tau_0(\mathbf{x})$, we take $\tau_0(\mathbf{x}) = \tau(\mathbf{x}_0)e^{-0.5(u^2+v^2)/\varepsilon^2}$, where $\varepsilon$ is an effective size of the small vicinity of $\mathbf{x}_0$ . Integral (3) can then be extended over the entire $(u,v)$ plane and can be found in explicit form:

$$-\omega\hat{u}_{\mathbf{x}_0}(\omega) = 2\pi\tau(\mathbf{x}_0)A[(\varepsilon^{-2} - i\omega a)(\varepsilon^{-2} - i\omega b)]^{-1/2} \propto |ab|^{-1/2}\omega^{-1}, \tag{9}$$

where $ab$ is equal to the curvature $K(x_0)$.

A rigorous proof can be found, e.g., in Fedoryuk (1987) (Ch.3, Theorem 2.1).

**2.3** Assume that $\mathbf{x}_0$ is an isolated point at the boundary of $\Sigma$ such that the isochrones $t_a(\mathbf{x}) = t_a(\mathbf{x}_0)$ is tangent to $\Sigma$ at $\mathbf{x}_0$ (the so-called *stationary point of the second kind*). Then the contribution $\hat{u}_{\mathbf{x}_0}(\omega)$ into the asymptotics of $\hat{u}(\omega)$ due to $\mathbf{x}_0$ is of order $O(\omega^{-2.5})$. In particular, if $t_a(\mathbf{x})$ is linear, then

$$|\hat{u}_{\mathbf{x}_0}(\omega)| = \omega^{-2.5}|\mathbf{d}|^{-1.5}|A\,\tau(\mathbf{x}_0)|\;|\;2\pi / K_{\partial\Sigma}(\mathbf{x}_0)\;|^{1/2}, \tag{10}$$

where $K_{\partial\Sigma}(\mathbf{x}_0)$ is the curvature of $\partial\Sigma$ at $\mathbf{x}_0$.

Stationary points of both types are well known in seismology and were extensively discussed by Bernard and Madariaga (1984), Spudich and Frazer (1984), and others.

A few words are in order to explain (10). We assume that the linear front is tangent to the boundary $\partial\Sigma$ at the point $\mathbf{x}_0 = \mathbf{0}$. If $t_a(\mathbf{x}) = cx''$, then the boundary near $\mathbf{x}_0$ is represented by the equation $x'' = kx'^2$. We choose a local coordinates $(x',z)$ such that $x'' = kx'^2 + z$. The contribution due to $\mathbf{x}_0 = \mathbf{0}$ is

$$-i\omega\hat{u}_{\mathbf{x}_0}(\omega) = A\int_{-\infty}^{\infty}dx\int_0^{\infty}dz\tau_0(x,z)e^{i\omega(kx^2+z)} \tag{11}$$

where $\tau_0(\mathbf{x}) = \tau(\mathbf{x})\varphi(\mathbf{x}-\mathbf{x}_0)$. Fitting $\tau_0(\cdot)$ by the function $\tau(\mathbf{x}_0)e^{-1/2x^2/\varepsilon^2}\psi(z)$, where $\psi$ is a finite smooth function with $\psi(0)=1$, we get the desired relation $-i\omega\hat{u}_{\mathbf{x}_0}(\omega) = O(\omega^{-3/2}), \omega >> 1$. This is easily seen, because we have dealt with the integral over $x$ when we derived (9), while the integral over $z$ is

$$\int_0^{\infty} e^{i\omega z}\psi(z)dz \approx i\omega^{-1}\psi(0), \omega >> 1 \tag{12}$$

(see (A5)). A rigorous proof of **2.3** can be found, e.g., in Fedoryuk (1987) (Ch.3, Th. 4.3).

**2.4** Assume that $\mathbf{x}_0$ is an angular point of the piece-wise smooth boundary $\partial\Sigma$ .Then $\partial\Sigma$ has two tangent half -lines at $\mathbf{x}_0$ If $\nabla t_a(\mathbf{x}_0)$ is not orthogonal to these lines, then the contribution into the asymptotics of $\hat{u}(\omega)$ due to $\mathbf{x}_0$ is of the order $O(\omega^{-3})$. This asymptotics is well known from the classical model where $\Sigma$ is a rectangle, $\tau(\mathbf{x})$ or $\Delta\dot{u}(\mathbf{x})$ is constant, and $t_r(\mathbf{x}) = (\mathbf{x}, \boldsymbol{\gamma}_r)/v$ (see e.g. Aki and Richards, 1980).

To explain the origin of the asymptotics, note that $t_a(\mathbf{x})$ is linear in a vicinity of $\mathbf{x}_0$.Choosing a suitable coordinate system with the origin at $\mathbf{x}_0$, the region $\Sigma$ can be considered to be a rectangular cone in a vicinity of $\mathbf{x}_0$. Then the contribution due to $\mathbf{x}_0$ will be given by

$$-i\omega\hat{u}_{\mathbf{x}_0}(\omega) = CA\tau(\mathbf{x}_0)\int_0^\infty \varphi(u)e^{i\omega au}du\int_0^\infty \phi(v)e^{i\omega bv}dv \tag{13}$$

where $\varphi$ и $\phi$ are arbitrary finite smooth functions with $\varphi(0) = \phi(0) = 1$, $(C, a, b)$ is defined by the linear approximation of $t_a(\mathbf{x})$ in a vicinity of $\mathbf{x}_0$ and by the local change of variables $\mathbf{x} \to (u, v)$.

By (12), the right-hand side of (13) is approximately $-CA\tau(\mathbf{x}_0)(ab)^{-1}\omega^{-2}$, $\omega >> 1$

As we can see, the class of smooth models of $(\tau(\mathbf{x}), \delta t_r(\mathbf{x}))$ is insufficient to discuss the $\omega^{-2}$ spectrum hypothesis. Therefore the Gusev (2014) model is of interest because it operates with the class of sufficiently rough functions of fractal type (see Fig.1). For an arbitrary function $f$ this can be expressed as follows:

$$|f(\mathbf{x}) - f(\mathbf{y})| \propto |\mathbf{x} - \mathbf{y}|^{H_f}, \quad 0 < H_f < 1 , \tag{14}$$

where the notation $a \propto b$ means that $c < a/b < C$ for some positive constants $c$ and $C$.

Such functions are difficult to construct and analyze. The situation becomes simpler, if we treat $f$ as a realization of a random field. In this case the fractal property of index $H_f$ means that

$$E|f(\mathbf{x}) - f(\mathbf{y})|^2 \propto |\mathbf{x} - \mathbf{y}|^{2H_f}, \quad 0 < H_f < 1, \tag{15}$$

where $E$ is the symbol of ensemble averaging.

To investigate how property (15) affects the HF asymptotics we shall study the r.m.s amplitude spectrum, viz.

$$r.m.s.\hat{u}(\omega) \doteq (E|\hat{u}(\omega)|^2)^{1/2}, \omega >> 1 . \tag{16}$$

Our final goals are :

(1) to characterize the asymptotics of $r.m.s.\hat{u}(\omega)$ in terms of fractal indexes $H_\tau$ and $H_r$ related to the source components $\tau(\mathbf{x})$ and $\delta t_r(\mathbf{x})$, respectively;

(2) to find conditions for the lack of the directivity in $r.m.s.\hat{u}(\omega)$ for high frequencies.

Note, that the stationary points of $t_a(\mathbf{x})$ that generate the $\omega^{-2}$ behavior of $\hat{u}(\omega)$ are impossible in the case of rough (no differentiable) $t_a(\mathbf{x})$. However, in the fractal case almost all points may be simultaneously a local extreme of $t_a(\mathbf{x})$ and intuitively we may expect the $\omega^{-2}$ behavior of $\hat{u}(\omega)$ at least for the quasi-regular case, $H_r \approx 1$ because any smooth function has $H_r = 1$

## 3. The Doubly Stochastic Model

**3.1** ***Specification of the model.*** Consider model (3) with stochastic independent components $\tau(\mathbf{x}), \delta t_r(\mathbf{x}), \mathbf{x} \in \Sigma$. Then (3) yields

$$\omega^2 E|\hat{u}(\omega)|^2 = A^2 \iint_{\Sigma\times\Sigma} m_\tau(\mathbf{x},\mathbf{y}) \exp(-i\omega((\mathbf{x}-\mathbf{y}),\mathbf{d})\varphi_\Delta(\omega|\mathbf{x},\mathbf{y}))\sigma(d\mathbf{x})\sigma(d\mathbf{y}) \ , \qquad (17)$$

where $m_\tau(\mathbf{x},\mathbf{y}) = E\tau(\mathbf{x})\tau(\mathbf{y})$ and

$$\varphi_\Delta(\omega|\mathbf{x},\mathbf{y}) = E\exp\{i\omega(\delta t_r(\mathbf{x}) - \delta t_r(\mathbf{y}))\}. \qquad (18)$$

To have an explicit form of (17), we consider two models:

$$\textbf{(A)}\ \delta t_r(\mathbf{x}) = \xi^2(\mathbf{x}), \qquad (19)$$

$$\textbf{(B)}\ \delta t_r(\mathbf{x}) = \xi(\mathbf{x}), \qquad (20)$$

which are based on a random field $\xi(\mathbf{x})$. We assume that

$(C_1)$ $\xi(\mathbf{x})$ *has zero mean, Gaussian two-dimensional distributions, and the fractal property (15) of index* $H_\xi$.

In virtue of the relation $\xi^2(\mathbf{x}) - \xi^2(\mathbf{y}) = (\xi(\mathbf{x}) + \xi(\mathbf{y}))(\xi(\mathbf{x}) - \xi(\mathbf{y}))$, the function $\xi^2(\mathbf{x})$ has locally the same index $H_\xi$ at the points where $\xi(\mathbf{x}) \neq 0$, i.e., the fractal indexes of $t_r(\mathbf{x})$ in models **A,B** can be treated as identical, $H_r = H_\xi$ We note that the natural condition $t_r(\mathbf{x}) \geq 0$ is violated in model **B.** However, both models are important to explain the directivity effect.

To specify the additional requirements to be imposed on the random functions $\xi(\mathbf{x})$ and $\tau(\mathbf{x})$, we introduce the following notation for a random functions

$$m_f(\mathbf{x},\mathbf{y}) = Ef(\mathbf{x})f(\mathbf{y}),\ B_f(\mathbf{x},\mathbf{y}) = \mathrm{cov}(f(\mathbf{x}), f(\mathbf{y})),\ \Delta_f^2(\mathbf{x},\mathbf{y}) = Var(f(\mathbf{x}) - f(\mathbf{y})). \qquad (21)$$

In addition, $\sigma_f^2(\mathbf{x}) = B_f(\mathbf{x},\mathbf{x})$

$(C_2)$ *All the functions listed in (21) for* $\xi(\mathbf{x})$ and $\tau(\mathbf{x})$ *are smooth enough* in $\Sigma\times\Sigma$ *out of the* diagonal $\mathbf{x}=\mathbf{y}$ ;

$\sigma_f^2(\mathbf{x})$ *for* $\xi(\mathbf{x})$ and $\tau(\mathbf{x})$ *is smooth and strictly positive on* $\Sigma$ *, including the boundary.*

$(C_3)$ *The fractal property (15) is specified as follows:*

$$\Delta_f(\mathbf{x},\mathbf{y}) = \alpha_f^2(\mathbf{x},\mathbf{y})|\mathbf{x}-\mathbf{y}|^{2H_f} + \beta_f(\mathbf{x},\mathbf{y}),\ \mathbf{x},\mathbf{y}\subset\Sigma, \tag{22}$$

*where* $\alpha_f(\mathbf{x},\mathbf{y}), \beta_f(\mathbf{x},\mathbf{y})$ *are smooth functions,* $\alpha_f^2(\mathbf{x}) \doteq \alpha_f^2(\mathbf{x},\mathbf{x}) > 0$*, and* $\beta_f(\mathbf{x},\mathbf{y}) = O(|\mathbf{x}-\mathbf{y}|^2)$.

$(C_4)$ *In addition, we assume*

$$\beta_f(\mathbf{x},\mathbf{y}) - (\sigma_f(\mathbf{x}) - \sigma_f(\mathbf{y}))^2 \neq O(|\mathbf{x}-\mathbf{y}|^3). \tag{23}$$

In other words, the left part of (23) is of order $O(|\mathbf{x}-\mathbf{y}|^2)$ but not $O(|\mathbf{x}-\mathbf{y}|^3)$. This is quite normal for smooth models of $\beta_f(\mathbf{x},\mathbf{y})$ and $\sigma_f(\mathbf{x})$.

The simplest example of a random field with properties $(C_2, C_3, C_4)$ is $f(\mathbf{x}) = a(\mathbf{x})w_H(\mathbf{x}) + b(\mathbf{x})$, where $a(\mathbf{x}) > 0, b(\mathbf{x})$ are smooth functions and $w_H(\mathbf{x})$ is the so-called Fractional Brownian Motion, i.e., a random field with zero mean and correlation function

$$Ew_H(\mathbf{x})w_H(\mathbf{y}) = 0.5(|\mathbf{x}|^{2H} + |\mathbf{y}|^{2H} - |\mathbf{x}-\mathbf{y}|^{2H}), 0 < H < 1. \tag{24}$$

In this case $\alpha_f^2(\mathbf{x},\mathbf{y}) = a(\mathbf{x})a(\mathbf{y})$ and $\beta_f(\mathbf{x},\mathbf{y}) = (b(\mathbf{x}) - b(\mathbf{y}))(a(\mathbf{x})|\mathbf{x}|^{2H} - a(\mathbf{y})|\mathbf{y}|^{2H})$. These functions are smooth if $0 \notin \Sigma$.

It is useful to mention the spectral scale of the signal smoothness. By definition, $f(\mathbf{x}), \mathbf{x}\in\Sigma$ has smoothness of the level $h_f$ if the signal spectrum $\hat{f}(\mathbf{k})$ has the asymptotics $\hat{f}(\mathbf{k}) \propto |\mathbf{k}|^{-1-h_f}, |\mathbf{k}|\to\infty$. In contrast to $H_f$, the index $h_f$ can be an arbitrary number. At the same time, small values of $h_f$ tell us nothing about the signal smoothness locally at a given point $\mathbf{x}$. If both indexes $H_f$ and $h_f$ exist simultaneously, then $h_f = H_f < 1$. Suppose that $h_f > 1$, then the signal $f(\mathbf{x})$ is differentiable. If $c < |\nabla f(\mathbf{x})| < C$, then $|f(\mathbf{x}) - f(\mathbf{y})| \propto |\mathbf{x}-\mathbf{y}|$, i.e. $H_f = 1$. For $h_f \le 0$ one has $I = \int_{|\mathbf{k}|>1} |\hat{f}(\mathbf{k})|^2 \sigma(d\mathbf{k}) = \infty$, i.e. the signal $f(\mathbf{x})$ does not exist as a function. Nevertheless, the case $-1 < h_f \le 0$ is discussed in applications and is well defined in the context of the generalized functions (Gel'fand and Shilov, 1964). In practice the spectrum power-law behavior is postulated in some wide but bounded range of $|k|$ only. As a result, the relation between $h_f$ and $H_f$ is not clear. Therefore any extrapolation of our results by $H_f$ based on the local relation $h_f = H_f$ is impossible without an additional analysis.

We now formulate and discuss the main results. To facilitate the reading, the proofs are relegated to the Appendix.

### 3.2 *Results.*

**Model A:** $\delta t_r(\mathbf{x}) = \xi^2(\mathbf{x})$. If properties $(C_1 - C_4)$ hold for $\xi(\mathbf{x})$ and $\tau(\mathbf{x})$ and $\Sigma$ is strictly convex area with smooth boundary, then the leading term of the $r.m.s.\hat{u}(\omega)$ asymptotics is formed by inner points $\dot{\Sigma}$ of $\Sigma$ and looks as follows

$$r.m.s.\hat{u}_{\dot{\Sigma}}(\omega) = C_{\tau\xi}|\mathbf{d}|^{-1+H_\xi/2}\omega^{-2-(1-H_\xi)/2}(1+O(|\omega\mathbf{d}|^{-\min(H_\xi,2H_\tau)})), \quad \omega \to \infty, \tag{25}$$

where $0 < H_\xi, H_\tau < 1$ are fractal indexes of $\xi(\mathbf{x})$ and $\tau(\mathbf{x})$ respectively,

$$C_{\tau\xi}^2 = 2\pi L(-H_\xi)\int_\Sigma m_\tau(\mathbf{x},\mathbf{x})\sigma_\xi(\mathbf{x})^{-1}\alpha_\xi^{-1}(\mathbf{x})\sigma(d\mathbf{x}), \tag{26}$$

$$L(h) = 2^{1+h}\Gamma(1+h/2)/\Gamma(-h/2) \approx -h, \quad h \to 0, \tag{27}$$

$\Gamma(h)$ is Gamma function, and $\alpha_\xi^2(\mathbf{x}) = \alpha_\xi^2(\mathbf{x},\mathbf{x})$ is given by (22).

The factor $C_{\tau\xi}^2$ does not depend on the spatial cross correlations of both $\tau(\mathbf{x})$ and $\delta t_r(\mathbf{x})$. Therefore (25) can be interpreted as a result of the incoherent radiation of the inner source points

(See Appendix **2** for more information and for the proof**).**

**The linear front:** $\delta t_r(\mathbf{x}) = 0$. If properties $(C_2, C_3)$ hold for $\tau(\mathbf{x})$, the inner points of $\Sigma$ yield the following contribution into the asymptotic of $r.m.s.\hat{u}(\omega)$:

$$r.m.s.\hat{u}_{\dot{\Sigma}}(\omega) \approx |A|C_\tau|\mathbf{d}|^{-1-H_\tau}\omega^{-2-H_\tau}(1+O(|\omega d|^{-2+2H_\tau})), \tag{28}$$

$$C_\tau^2 = -\pi L(2H_\tau)\int_\Sigma \alpha_\tau^2(\mathbf{x})\sigma(d\mathbf{x}) \; . \tag{29}$$

The structure of $C_\tau^2$ show, that the contribution (28) is formed by the incoherent radiation of the inner source points of $\Sigma$.

Suppose that $\partial\Sigma$ is a smooth boundary of nonzero curvature at each point and $H_\tau > 1/2$. Then the leading term of the $r.m.s.\hat{u}(\omega)$ asymptotics comes from the boundary:

$$r.m.s.\hat{u}_{\partial\Sigma}(\omega) \approx |A|\widetilde{C}_\tau|\mathbf{d}|^{-1.5}\omega^{-2.5}, \tag{30}$$

where

$$\widetilde{C}_\tau^2 = 2\pi[m_\tau(\mathbf{x}_+,\mathbf{x}_+)\kappa_+^{-1} + m_\tau(\mathbf{x}_-,\mathbf{x}_-)\ \kappa_-^{-1} - 2sm_\tau(\mathbf{x}_+,\mathbf{x}_-)\sin(\omega(\mathbf{x}-\mathbf{x},\mathbf{d}))(\kappa_+\kappa_-)^{-1/2}], \qquad (31)$$

$\mathbf{x}_\pm$ are boundary points, at which $\mathbf{d}$ is orthogonal to $\partial\Sigma$, $\kappa_\pm$ is the curvature of $\partial\Sigma$ at $\mathbf{x}_\pm$, $s = \mathrm{sgn}(\mathbf{d},\mathbf{n}_+)$ and $\mathbf{n}_+$ is the outer normal to $\partial\Sigma$ at $\mathbf{x}_+$.

The extreme values of $\widetilde{C}_\tau^2$ are $C_\pm^2 = 2\pi E[\tau(\mathbf{x}_+)\kappa_+^{-1/2} \pm \tau(\mathbf{x}_-)\kappa_-^{-1/2}]^2$. If $\tau(\mathbf{x}_+)$ and $\tau(\mathbf{x}_-)$ are correlated, we can consider (30) as the result of coherent radiation of the boundary source points.

For the proof see Appendix **3**.

**Model B:** $\delta t_r(\mathbf{x}) = \xi(\mathbf{x})$. Assume that $\xi(\mathbf{x})$ has fractal property of the following type:

$$\lim_{y\to x} \Delta_\xi(\mathbf{x},\mathbf{y})/|\mathbf{x}-\mathbf{y}|^{2H_\xi} = \alpha_\xi^2(\mathbf{x}) \ , \ \mathbf{x},\mathbf{y} \subset \Sigma, \qquad (32)$$

where $\alpha_\xi(\mathbf{x})$ is a smooth positive function and $0 < H_\xi < 1$. Then under conditions $(C_1, C_2)$ we have

$$r.m.s.\hat{u}(\omega) \approx |A| C_{\tau\xi}\omega^{-2-(1-H_r)/H_r}, \quad \omega \to \infty, \qquad (33)$$

where

$$C_{\tau\xi}^2 = 2^{1/H_\xi}\pi\Gamma(1+H_\xi^{-1})\int_\Sigma m_\tau(\mathbf{x},\mathbf{x})\alpha_\xi^{-2/H_\xi}(\mathbf{x})\sigma(d\mathbf{x}). \qquad (34)$$

The amplitude factor (34) is formed by the incoherent radiation of the inner source points of $\Sigma$.

Assume that $\xi(\mathbf{x}), \mathbf{x} = (x', x'')$ has different smoothness along different axes such that

$$c_-^2 < \Delta_f^2(\mathbf{x},\mathbf{y})/(|x'-y'|^{2H'} + |x''-y''|^{2H''}) < c_+^2. \qquad (35)$$

Then $r.m.s.\hat{u}(\omega) = O(\omega^{-2-(1-\overline{H}_r)/\overline{H}_r})$, where $\overline{H}_r = 2/(1/H' + 1/H'')$ is the enharmonic mean of the indexes.

In the both cases the resulting asymptotic relations are independent on the directivity $\mathbf{d}$.

For the proof see Appendix **A4.**

## 3.3 Comments.

***Order of the HF asymptotics***. The stochastic models with fractal components $\tau(\mathbf{x})$ and $\delta t_r(\mathbf{x})$ discussed here show a power-law asymptotics in the displacement spectra, more precisely, $r.m.s.\hat{u}(\omega) \approx C\omega^{-\theta}, \omega \to \infty$.

Naturally, that there is no universal relationship of the exponent $\theta$ to the fractal indexes $H_\tau$ and $H_r$. However, the models with $\delta t_r(\mathbf{x}) \neq 0$ behave in the same manner in relation to the $\omega^{-2}$ hypothesis. Recall that

$$\theta = 2 + (1 - H_r)/2 \text{ (Model A) and } \theta = 2 + (1 - H_r)/H_r \text{ (Model B)}. \qquad (36)$$

One has

(a) $\theta > 2$ , i.e., the signal $u(t)$ is smoother than that observed;

(b) $\theta \to 2$, when $H_r \to 1$, i.e., when the running rupture front becomes quasi-regular (not a fractal line).

These facts are nontrivial. As we know, in the case $H_r = 1$ the observed asymptotics $O(\omega^{-2})$ is possible due to the stationary points of the arrival time function $t_a(\mathbf{x})$ (see 2.2). However, the stationary points are unstable and disappear as soon as $H_r < 1$. The results show that the effect of the stationary points almost remains for $H_r \sim 1$ only; moreover, a slight fractality of $t_a(\mathbf{x})$ stabilizes the quasi $\omega$-square behavior. For this reason, of the two models $t_a(\mathbf{x})$ shown in Fig1(a, b), the model with the smoother isochrones (Fig. 1b) would be preferred from our theoretical background (the opposite choice has been made by Gusev(2014)).

(c) the role of the stress drop component in producing the HF spectrum is minimal both for the smooth and for the fractal cases of $\tau(\mathbf{x})$: the effect of $\tau(\mathbf{x})$ is observed in the amplitude of the spectral asymptotics only (see 26, 34).

(d) $\theta$ is a decreasing function of $H_r$ (see (36)).

To clarify the generality of property (d), we assume $\tau(\mathbf{x}) = 1$ for simplicity, because $\theta$ does not depend on $\tau(\mathbf{x})$. Then

$$iA^{-1}\omega\hat{u}(\omega) = \int_\Sigma \exp(i\omega t_a(\mathbf{x}))\sigma(d\mathbf{x}) = \int_{t_-}^{t_+} e^{i\omega t} dL_a(t), \quad (37)$$

where

$$L_a(t) = \int_\Sigma \chi(t - t_a(\mathbf{x}))\sigma(d\mathbf{x}) \quad (38)$$

is the measure of $\mathbf{x}$-points related to the isochrones with arrival time $t_a < t$; $t_-$ and $t_+$ are initial and final arrival times at the receiver site, respectively. If $L_a(t)$ has density $l(t) = dL_a(t)/dt$, we can interpret $l(t)$ as a *rate of the isochrones*. Note, that $l(t)$ is sometimes used to interpret the ground acceleration (see e.g., Burridge (1963), Spudich and Fraser (1984)).

Relation (37) means that the exponent in the power-law asymptotics of $\hat{u}(\omega)$ is closely related to the smoothness of $l(t)$ extended by $0$ outside of $(t_-, t_+)$. In the smooth case of $t_a(x)$, the measure $dL_a(t)$ has a piecewise smooth density $l(t)$. Any isolated stationary point of $t_a(\mathbf{x})$ produces a jump in $l(t)$. For example, suppose the arrival time function has a local maximum in a vicinity of $\mathbf{x}_0$, namely $t_a(\mathbf{x}) = t_0 - \mu|\mathbf{x} - \mathbf{x}_0|^2$. By (38), the contribution of this vicinity in $dL_a(t)$ is $d(\pi(t_0 - t)_+ / \mu)$, where $(a)_+ = \max(a,0)$. Hence $l(t)$ has

the jump $-\pi/\mu$ at $t=t_0$. Any finite jump of $l(t)$ causes the $\omega^{-1}$ - HF behavior of $\hat{\dot{u}}(\omega)$, i.e., $\omega^{-2}$ asymptotics of $\hat{u}(\omega)$. This is a special case of the statement (8).

Consider a more general example of an isolated local extreme at point $\mathbf{x}_0$, namely,

$$t_a(\mathbf{x}) = t_0 - \mu|\mathbf{x}-\mathbf{x}_0|^H,\ H>0\ , \tag{39}$$

where $\mu \neq 0$. In the case $0<H<1$, $x_0$ is a singular point of $t_a(\mathbf{x})$ of the index $H$, and in the case $H>1$, $x_0$ resembles a stationary point because $\nabla t_a(\mathbf{x}_0)=0$ or $\nabla \delta t_r(\mathbf{x}_0)=\mathbf{d}$. The last relation means that a local extreme of type (39) is unstable for any $H>1$.

By definition of the $dL_a(t)$ measure, the contribution of $\mathbf{x}_0$ in $l(t)$ is

$$l_0(t) = -\pi\mu^{-1}\cdot(2/H)(\mu^{-1}(t_0-t))_+^{2/H-1}. \tag{40}$$

According to (A5), the contribution of $t_0$ in the HF asymptotics of the Fourier transform of $l_0(t)$ looks as follows:

$$\left|\hat{l}_0(\omega)\right| = \pi|\mu\omega|^{-2/H}\Gamma(1+2/H). \tag{41}$$

Consequently, the contribution of $x_0$ in the spectrum $\hat{u}(\omega)$ is $O(\omega^{-1-2/H})$. The exponent of this asymptotics $\theta_0 = 1+2/H$ is a decreasing function of $H$. Considering the fractal models of $t_a(\mathbf{x})$ as a superposition of the singularities of type (40) with fixed $H=H_r$, the fact that $\theta$ is a decreasing function of $H_r$ looks natural. However, the fractal models of $t_a(\mathbf{x})$ are more complicated and we have $\theta<\theta_0$ instead of $\theta=\theta_0$.

Along the way, the example (40) shows that the smooth model of $t_a(\mathbf{x})$ allows us to get a spectral asymptotics with any exponent $\theta>1$.

***The case*** $\delta t_r(\mathbf{x})=0$. Only in this case the stress drop component affects $\theta$ ($\theta=2+H_\tau$ ). The desirable asymptotics with $\theta\approx 2$ is generated by $\tau(\mathbf{x})$ with $H_\tau \ll 1$, i.e., when the stress drop function $\tau(\mathbf{x})$ is as rough as possible. This is natural, because

$$iA^{-1}\omega\hat{u}(\omega) = |d|^{-1}\int_{t_-}^{t_+} e^{i\omega t}\tilde{\tau}(t)dt\,, \tag{42}$$

where $\tilde{\tau}(t)$ is the integral of $\tau(\mathbf{x})$ along the linear isochrones $t_a(\mathbf{x})=t$; given (22), $\tilde{\tau}(t)$ is more smooth compared with the cross-sections of $\tau(\mathbf{x})$ (Andrews (1981) notes this fact for self-affine random fields). Therefore the smoothness of $\tilde{\tau}(t)$ increases with $H_\tau$.

In fact, Andrews (1981) and Herrero, Bernard (1994) considered the case $\delta t_r(\mathbf{x}) = 0$ using a special model of $\tau(\mathbf{x})$ (or $\Delta\dot{u}(\mathbf{x})$). The model was a random field with the power spectrum $f_0(\mathbf{k}) = |\mathbf{k}|^{-2-2h}, h \ge 0$. Because the spectrum is integrable over $|\mathbf{k}| > 1$ for $h > 0$ only, $\tau(\mathbf{x})$ does not exist as a function, unless $h > 0$. Therefore, it is useful to dwell on this example to underline the importance of the source boundary in the HF radiation.

Assume that $0 < h < 1$. Then there is a random continuous function $\tau(\mathbf{x})$ with zero mean and the power spectrum $f_0(\mathbf{k})$, such that

$$E|\tau(\mathbf{x}) - \tau(\mathbf{y})|^2 = \alpha_\tau^2 |\mathbf{x} - \mathbf{y}|^{2h}, \tag{43}$$

where $\alpha_\tau$ is constant (see (24)). Fourier transform of the generalized function $f_0(\mathbf{k})$ (Gel'fand and Shilov, 1964) can be used to show that $\alpha_\tau^2 L(2h) = -2\pi$ (see (27) for $L(\cdot)$). Comparing (43) with (22), one has $H_\tau = h$. Therefore we can use the HF asymptotics (28-30). Due to relation $\alpha_\tau^2 L(2h) = const$, the amplitude in the asymptotics (28, 29) is independent of $H_\tau$. Therefore it is formally meaningful for $H_\tau = 0$. At the same time, the asymptotics (30) related to $\partial\Sigma$ involves an unbounded factor $\tilde{C}_\tau$ at $H_\tau = 0$. Indeed, $\tilde{C}_\tau^2$ linearly depends on $E\tau^2(\mathbf{x}) = \sigma_\tau^2(\mathbf{x}), \mathbf{x} \in \partial\Sigma$. By (43), $E\tau^2(\mathbf{x}) \propto \alpha_\tau^2 \sim \pi / H_\tau \to \infty$ as $H_\tau \to 0$. Hence, the passage to the limit $(H_\tau \to 0)$ in (28) is incorrect. We can see that the boundary component of the HF spectral asymptotics (30) is important not only for the case $H_\tau > 1/2$.

***Directivity effect.***

The results for models **A** and **B** are directly opposite in relation to directivity: the effect is present in **A** and is lacking in **B.** The type of fractality of $t_a(\mathbf{x})$: isotropic (15) or anisotropic (35) is irrelevant to this result.

The key to the understanding of this fact is contained in the main relation (17):

$$\omega^2 E|\hat{u}(\omega)|^2 = A^2 \iint_{\Sigma\times\Sigma} m_\tau(\mathbf{x},\mathbf{y}) \exp(-i\omega((\mathbf{x}-\mathbf{y}),\mathbf{d})\varphi_\Delta(\omega|\mathbf{x},\mathbf{y}))\sigma(d\mathbf{x})\sigma(d\mathbf{y}), \tag{44}$$

where $\varphi_\Delta(\omega|\mathbf{x},\mathbf{y})$ is the Fourier transform of the distribution of random variable $\delta t_r(\mathbf{x}) - \delta t_r(\mathbf{y})$. Suppose that $\varphi_\Delta(\omega|\mathbf{x},\mathbf{y})$ decays over frequency very fast outside the diagonal $\mathbf{x} - \mathbf{y} = \mathbf{0}$; in this case the asymptotics of the integral is controlled by a neighborhood of that diagonal only. In turn, the oscillating term $\exp(-i\omega((\mathbf{x}-\mathbf{y}),\mathbf{d})$ is near to 1 at the diagonal, hence the asymptotics no longer depends on the directivity vector $\mathbf{d}$.

It is known that the Fourier transform of a finite smooth function decays faster than any power-law function, $|\omega|^{-N}$ (Gel'fand and Shilov, 1964). For this reason, to have the desirable property of $\varphi_\Delta(\omega|\mathbf{x},\mathbf{y})$ one has to require that the densities of 2-D distributions of $\delta t_r(\mathbf{x})$ should be finite and sufficiently smooth. By physical reasons, $|\delta t_r(\mathbf{x})|$ is bounded. For convenience of calculation we can relax this condition and consider distributions with rapidly decreasing "tails".

The above properties of 2-D distributions (smoothness and light tails) can be automatically transferred to one-dimensional distributions of $\delta t_r(\mathbf{x})$ .Therefore it is useful to have a look at the 1-D distributions of $\delta t_r(\mathbf{x})$ in the above models. By (19, 20), $\delta t_r(\mathbf{x})$ is equal to $\xi^2(\mathbf{x})$ and $\xi(\mathbf{x})$ in models **A** and **B,** respectively**,** where $\xi(\mathbf{x})$ is a Gaussian random variable with standard deviation $\sigma = \sigma_\xi(\mathbf{x})$. Therefore the distribution density, $p(t)$ of $\delta t_r(\mathbf{x})$ is

$$p(t) = (2\pi t\sigma^2)^{-1/2} \exp(-t\sigma^{-2}/2),\ t > 0 \text{ (model } \mathbf{A}) \tag{45}$$

and

$$p(t) = (2\pi\sigma^2)^{-1/2} \exp(-t^2\sigma^{-2}/2),\ |t| < \infty \text{ (model } \mathbf{B}). \tag{46}$$

Model **A** has discontinuous density at t=0, and therefore produces the directivity effect (see above).

Gusev uses the Rayleigh law for $\delta t_r(\mathbf{x})$ ,

$$p(t) = t\sigma^{-2} \exp(-t/\sigma), t \geq 0 \text{ (Gusev (2014), section 5).} \tag{47}$$

This density has a discontinuous second derivative at 0. As a result, the reduction of the directivity effect in Gusev's computer simulations "looks less convincing than it actually is".

The lack of directivity in the framework of the doubly stochastic earthquake source model is unstable, because it is based on the smoothness of 2-D distributions of $\delta t_r(\mathbf{x})$ . The above arguments do not use a specific form of the distribution of $\delta t_r(\mathbf{x}) - \delta t_r(\mathbf{y})$ . Therefore the Gaussian property used in the models **A** and **B** is not critical to our conclusions.

### 3.4 Crack nucleation point: contribution in the spectrum asymptotics.

It is a thought that the crack starts from an inner point of $\Sigma$ and propagates in radial directions with a constant velocity (Aki and Richards, 1980). Therefore we may improve the previous doubly stochastic earthquake source models by adding a circular dominant front:

$$t_r(\mathbf{x}) = |\mathbf{x} - \mathbf{x}_0|/v + \delta t_r(\mathbf{x}) \ , \qquad v < c, \tag{48}$$

at the initial stage of the seismic event. From the mathematical point of view this means that the model acquires a new singular point, namely $\mathbf{x}_0$. Hence we have to find the contribution of this point $\hat{u}_{\mathbf{x}_0}(\omega)$ in the HF asymptotics of the signal. To do this, we use the following relation:

$$-i\omega\hat{u}_{\mathbf{x}_0}(\omega) = Ae^{it_0\omega}\int_\Sigma \varphi(\mathbf{x})\tau(\mathbf{x})\exp(i\omega|\mathbf{x}|v^{-1} - i\omega(\mathbf{x},\boldsymbol{\gamma})c^{-1} + i\omega\delta t_r(\mathbf{x}))\sigma(d\mathbf{x}) \ . \tag{49}$$

Here $\varphi$ is an arbitrary smooth function such that $\varphi(\mathbf{x}) = 1$ near $\mathbf{x}_0$ and $\varphi(\mathbf{x}) = 0$ outside of small vicinity of $\mathbf{x}_0$. A special case of (49) with a smooth function $\varphi\tau$ and $\delta t_r(x) = 0$ arises in the Savage model (see Aki and Richards, 1980, Ch. 14-16) implying $\hat{u}_{\mathbf{x}_0}(\omega) = O(\omega^{-3})$. This fact holds for any smooth function $\tau(\mathbf{x})$. More generally, if $\tau(\mathbf{x})$ has the fractal property of index $0 < H_\tau < 1$, then $r.m.s.\hat{u}_{\mathbf{x}_0}(\omega) = O(\omega^{-2-H_\tau})$. It is important, that the joining of the nucleation point to any stochastic model considered above does not change the order of the signal HF asymptotics. More precisely

$$r.m.s.\hat{u}_{\mathbf{x}_0}(\omega) = O(\omega^{-\theta_{\mathrm{mod}}})\|\varphi\| \tag{50}$$

where the exponent $\theta_{\mathrm{mod}}$ is given by (25, 28 and 33) for the models**: A, LF** (Linear Front), **B** respectively. Since $\varphi$, $\varphi(\mathbf{x}_0) = 1$, can be selected with arbitrary small norm $\|\varphi\| = (\int \varphi^2(\mathbf{x})\sigma(d\mathbf{x}))^{1/2}$, the role of the nucleation point in the global HF radiation is small. (See Appendix 5 for the proof).

**3.5. Numerical analysis**

The theoretical results (25-35) need confirmation in the frequency range of practical interest 1-10Hz. This can be done by numerical analysis of $\hat{u}(\omega)$. Below such an analysis is based on Fig 2 kindly prepared by A.Gusev.

Following to Gusev (2014), the source models are specified as follows: S-wave velocity is $c = 3.5$ km/s; mean velocity of rupture front is $v = 2.625$ km/s; the fault patch is the rectangle $L \times W$ with $L = 2W = 33.6$ km; the size of the source corresponds approximately to the magnitude range M=6.7-6.8. Because the linear component $(\mathbf{x}, \boldsymbol{\gamma}_r)/v$ of $t_r(\mathbf{x})$ dominates the rupture propagation model, the following normalization is used:

$$\sigma(\delta t_r)/(Lv^{-1}) = 0.2, \tag{51}$$

where $\sigma(\cdot)$ is the standard deviation of the random field. Roughly speaking, (51) means that the variation coefficient of $t_r(\mathbf{x})$ is small.

Gusev (2014) generates $\xi(\mathbf{x}), \tau(\mathbf{x})$ as a homogeneous Gaussian field with discrete spectrum $c_H|\mathbf{k}|^{-2-2H}$, where $\mathbf{k} = (i, j) \neq 0$ is integer, $-N/2 < i, j \leq N/2$, $\sigma(\xi(\mathbf{x})) = 1$, and $H = H_r$, $H_\tau$ respectively.

The source functions are independent and defined on lattice of side $\delta = L / N$ . That is, $N$ is responsible both for the space discretization and for fidelity of the fractal properties of the source components in the $\mathbf{k}$ -space. The value $N = 1024$ that we are using is not too large in the both case. Therefore the numerical results are largely illustrative.

In the case of linear front, $\delta t_r(\mathbf{x}) = 0$ the space discretization problem is not so critical, owing to linearity of isochrones (see above). The theoretical power-law asymptotics in this case,

$$r.m.s.\hat{u}(\omega) \approx C \; |\mathbf{d}|^{-\theta_d} \omega^{-\tilde{\theta}}, \quad \omega \to \infty, \tag{52}$$

involves $\theta_d = 1 + H_\tau$ and $\tilde{\theta} = H_\tau$ . Figure 2 shows that $\tilde{\theta}$ is in a good agreement with its numerical estimates in the range 1-10 Hz for $H_\tau = 0.2$ (Fig. 2a) and $H_\tau = 0.8$ (Fig.2b). The boundary effect in $\tilde{\theta}$ for $H_\tau = 0.8$ is small because of the geometry of $\partial\Sigma$ .

In the case $\delta t_r(x) \neq 0$ the effect of the space discretization is more essential for the spectral asymptotics. Fig 2(c, e) represent the model **A** with $H_r = 0.2$ (c) and $H_r = 0.8$ (e). The numerical estimations of $\tilde{\theta}$ vs. the theoretical one for frequency range 1-3Hz are 0.2 vs. 0.4(c) and 0 vs. 0.1(e).

Looking at Fig 2(c, d) we can see a parasitic fragment in spectrum $\hat{u}(\omega)$ in the range $\omega \geq 3$ Hz, it increases linearly and distorts $\hat{u}(\omega)$ . The following explains this fact and predict unite slope (in log-log scale) in the parasitic spectrum.

A discrete analogue of (17) for $E\left|\hat{u}(\omega)\right|^2$ looks as follows:

$$(17)_{discr.} = (\sum\nolimits_{l=j} + \sum\nolimits_{l \neq j}) A^2 m_\tau(\mathbf{x}_l, \mathbf{x}_j) \exp(-i\omega((\mathbf{x}_l - \mathbf{x}_j), \mathbf{d}) \varphi_\Delta(\omega|\mathbf{x}_1 - \mathbf{x}_j))\delta^4, \tag{53}$$

where $\delta$ is the lattice spacing. The first sum is approximately $\tilde{A}^2\delta^2$ , where $\tilde{A}^2 = A^2 \int_\Sigma E\tau^2(\mathbf{x})\sigma(d\mathbf{x})$ , while the second sum is a correct approximation of $E\left|\hat{u}(\omega)\right|^2$ . Hence, if $\hat{u}(\omega) \to 0$ as $\omega \to \infty$ , then the numerical estimate of $r.m.s.\hat{u}(\omega)$ is $\tilde{A}\delta \cdot \omega + o(\omega)$ . The actual procedure used to estimate $r.m.s.\hat{u}(\omega)$ is different, but the origin of the parasitic spectrum remains the same. Fig 2 shows that the range of $\omega$ , where the linear parasitic spectrum $\tilde{A}\delta\omega$ dominates, depends of the source model parameters.

Fig 2(d, f) represent model **B** with $H_r = 0.2$ and $0.8$ , respectively. In both cases the theory predicts lack of directivity. To control the directivity effect, the spectrums on Fig 2 depend on the angle $\alpha$ between the direction of rupture, $\boldsymbol{\gamma}_r$ and the recording station, $\boldsymbol{\gamma}$ . To be specific, we use $\alpha = 0^o, 90^o, 180^o$ . Fig 2d (but not Fig2f) supports the spectrum independence on $\alpha$ in the range $\omega: 1-3$ Hz.

To clarify this "contradiction", let us accept the theoretical result for Fig 2d in the frequency range near $\omega = 2$ Hz. As it follows from Section 3.3(*Directivity effect*), the spectrum asymptotics is independent of $\alpha$ , if

$$\varphi_\Delta(\omega|\mathbf{x},\mathbf{y}) = Ee^{i\omega(\delta t_r(\mathbf{x})-\delta t_r(\mathbf{y}))}, \qquad \omega > \omega_0 \tag{54}$$

is small enough outside of a small neighborhood of the diagonal $\mathbf{x} = \mathbf{y}$ . For the model of $\delta t_r(\mathbf{x})$ we are using this condition roughly means that $\omega_0 |\mathbf{x}-\mathbf{y}|^{H_r} > 1$. In case of Fig2d, one has $H_r = 0.2$ and $\omega_0 = 2$ Hz. Using $H_r = 0.8$ , the corresponding $\omega_0$ has to be $2^{0.8/0.2} = 16$ Hz. This frequency belongs to the range where the parasitic spectrum in Fig2f is dominant. These examples show the difficulties of the numerical analysis both of the $\omega^{-2}$ hypothesis and of the directivity suppression effect

## 4. Conclusions

We used the doubly stochastic earthquake source model in order to explain the $\omega^{-2}$ behavior of the displacement spectra for the far-field body waves. The model involves two components of the fractal type: the local stress drop over the fault, $\tau(\mathbf{x})$ and the arrival time function $t_a(\mathbf{x})$. The fractality is controlled by the indices of smoothness $H_\tau$ and $H_r$ , respectively. In the class of doubly differentiable functions of $t_a(\mathbf{x})$ it is possible to get the $\omega^{-\theta}$ spectral behavior with any index $\theta > 1$. However, the asymptotics with $\theta \le 2$ is unstable under small perturbations of $t_a(\mathbf{x})$ and receiver site location. This results from the instability of extreme points of $t_a(\mathbf{x})$.

The fractal models of $t_a(\mathbf{x})$ where almost any point is extreme, allow us to get a stable $\omega^{-\theta}$ asymptotics with any $\theta > 2$ . In particular, one can get the quasi-$\omega^{-2}$ behavior of the spectra using $t_a(\mathbf{x})$ fields of an almost regular type $(H_r \approx 1)$. What is important is that the order of the spectrum asymptotics depends on the stress drop component if only the rupture front is linear. In this case the desirable quasi-$\omega^{-2}$ behavior is generated by the stress drop function $\tau(\mathbf{x})$ of minimal smoothness $H_\tau << 1$ .

In the fractal stochastic models the amplitude factor of the HF asymptotics is stable and usually is formed by the incoherent radiation of the inner source points.

The directivity effect does take place, but not for all fractal models. In our analysis this is true for model **B** only, even without imposing the additional requirement that the index of smoothness of the arrival function is identical in all directions. It thus appears that the isotropic fractal property of $t_a(\mathbf{x})$ is not very important for reducing the directivity effect. As it turns out, the degree of smoothness of 2-D distributions of $t_a(\mathbf{x})$ is the main controlling factor for the directivity. For this reason the lack of directivity property in the framework of the doubly stochastic earthquake source model is unstable. Moreover, it is not clear how to reformulate this ensemble property in physical terms. Nevertheless, the instability is in agreement with the opposite expert conclusions related to statistical significance of the rupture directivity effect on different faults (see papers by Calderoni et al (2014) and Kurzon et al (2014) with observational evidences of the effect).

## Acknowledgements

A.Gusev called my attention to the issues considered here and kindly made the numerical calculations. Our discussions have been highly productive for this work. Y. Ben-Zion has informed me of the results that support the rupture directivity effect. Comments and numerous suggestions of two anonymous reviewers considerably affected the final text of this paper.

# Appendix

## A1. Auxiliary statements

***Shivakumar and Wang*** (1979). Consider

$$I(\mathbf{k}|\phi) = (2\pi)^{-1} \int_{R^2} \phi(\mathbf{z}) e^{i(\mathbf{z},\mathbf{k})} \sigma(d\mathbf{z}), \tag{A1}$$

where $\phi(\mathbf{z})$, $\mathbf{z} = (z_1, z_2)$ is a finite smooth function in $R^2 \setminus \{0\}$ such that,

$$\phi(\mathbf{z}) = z_1^{n_1} z_2^{n_2} |\mathbf{z}|^h + \varphi(\mathbf{z}), \tag{A2}$$

$n_i \geq 0$ are integer, $h$ is a real number. Assume that $h + n_1 + n_2 \leq q$, $2m - 3 \leq q < 2m - 1$, $m \geq 0$ is integer, and

$$(\Delta^j \varphi)(\mathbf{z}) = \begin{cases} O(|z|^{q-2j+1}), j = 0,...m, q \neq 2m-3 \\ O(|z|^{q-2j+2}), j = 0,...m, q = 2m-3 \end{cases}, \quad |z| \to 0. \tag{A3}$$

Then for $h \neq -2, -4, -6, ...$

. $$I(\mathbf{k}|\phi) = L(h)(-i\partial/\partial k_1)^{n_1}(-i\partial/\partial k_2)^{n_2}|\mathbf{k}|^{-(2+h)} + \delta_m(\mathbf{k}), \quad |\mathbf{k}| >> 1 \quad , \tag{A4}$$

where the error term is $\delta_m(\mathbf{k}) = (-1)^m |\mathbf{k}|^{-2m} I(k|\Delta^m \varphi) = o(|\mathbf{k}|^{-2m})$ and $L(h)$ is given by (27).

***Erdelyi*** *(1955)*. If $\phi(t) = t^{\beta-1} f(t), \beta > 0$ is a smooth function on semi-axis $t > 0$ and $\phi(t) = 0$ for $t > a > 0$, then

$$\int_0^\infty e^{i\omega t}\phi(t)dt = \omega^{-\beta} f(0)\Gamma(\beta)e^{i\pi\beta/2} + \delta(\omega), \omega \to \infty \tag{A5}$$

where $|\delta(\omega)| \le \int_0^\infty t^{\beta-1}|f'(t)|dt/\omega$, if $0 < \beta < 1$.

## A2. Model A.

In this model $\delta t_r(\mathbf{x}) = \xi^2(\mathbf{x})$, $(\xi(\mathbf{x}), \xi(\mathbf{y}))$ is Gaussian vector with zero mean and the correlation matrix

$$\mathbf{B}_\xi = \begin{pmatrix} \sigma_\xi^2(\mathbf{x}) & B_\xi(\mathbf{x},\mathbf{y}) \\ B_\xi(\mathbf{x},\mathbf{y}) & \sigma_\xi^2(\mathbf{y}) \end{pmatrix} \tag{A6}$$

such that $\det \mathbf{B}_\xi > 0$ if $\mathbf{x} \ne \mathbf{y}$ (see (21) for notation). In other words, the distribution density of $(\xi(\mathbf{x}), \xi(\mathbf{y}))$ is

$$p_{\mathbf{B}_\xi}(\mathbf{u}) = (2\pi \det \mathbf{B}_\xi)^{-1} \exp(-0.5(\mathbf{u}, \mathbf{B}_\xi^{-1}\mathbf{u}), \ \mathbf{u} = (u_1, u_2). \tag{A7}$$

Therefore

$$\varphi_\Delta(\omega|\mathbf{x},\mathbf{y}) = E\exp\{i\omega(\xi^2(\mathbf{x}) - \xi^2(\mathbf{y})\}$$

$$= (2\pi)^{-1}(\det \mathbf{B}_\xi)^{-1/2}\iint \exp(-0.5(\mathbf{u}, \mathbf{B}_\xi^{-1}\mathbf{u}) + i\omega(u_1^2 - u_2^2))du_1 du_2. \tag{A8}$$

Using notation

$$\widetilde{\mathbf{B}}_\xi^{-1} = \mathbf{B}_\xi^{-1} - \begin{pmatrix} 2i\omega & 0 \\ 0 & -2i\omega \end{pmatrix}, \tag{A9}$$

and the relation $\int p_{\mathbf{B}}(\mathbf{u})du_1 du_2 = 1$, one get

$$\varphi_\Delta(\omega|\mathbf{x},\mathbf{y}) = (\det \mathbf{B}_\xi \widetilde{\mathbf{B}}_\xi^{-1})^{-1/2} = [\omega^2 b_\xi^2(\mathbf{x},\mathbf{y}) - 2i\omega(\sigma_\xi^2(\mathbf{x}) - \sigma_\xi^2(\mathbf{y})) + 1]^{-1/2}$$

$$= \omega^{-1}[b_\xi^2(\mathbf{x},\mathbf{y}) + O(\omega^{-1})]^{-1/2}, \tag{A10}$$

where

$$b_\xi^2(\mathbf{x},\mathbf{y}) = 4\det \mathbf{B}_\xi = 2(\sigma_\xi^2(\mathbf{x}) + \sigma_\xi^2(\mathbf{y}))\Delta_\xi^2(\mathbf{x},\mathbf{y}) - (\sigma_\xi^2(\mathbf{x}) - \sigma_\xi^2(\mathbf{y}))^2 - \Delta_\xi^4(\mathbf{x},\mathbf{y}). \tag{A11}$$

By condition $(C3)$,

$$\Delta_\xi^2(\mathbf{x},\mathbf{y}) = E|\xi(\mathbf{x})-\xi(\mathbf{y})|^2 = \alpha_\xi(\mathbf{x},\mathbf{y})|\mathbf{x}-\mathbf{y}|^{2H_\xi} + \beta_\xi(\mathbf{x},\mathbf{y}), \tag{A12}$$

where $\beta_\xi(\mathbf{x},\mathbf{y}) = O(|\mathbf{x}-\mathbf{y}|^2)$ . In addition,

$$m_\tau(\mathbf{x},\mathbf{y}) = E\tau(\mathbf{x})\tau(\mathbf{y}) = m_\tau(\mathbf{x})m_\tau(\mathbf{y}) + (\sigma_\tau^2(\mathbf{x})+\sigma_\tau^2(\mathbf{y}))/2 - \Delta_\tau^2(\mathbf{x},\mathbf{y})/2$$

$$= \mu_\tau(\mathbf{x},\mathbf{y}) - \alpha_\tau^2(\mathbf{x},\mathbf{y})|\mathbf{x}-\mathbf{y}|^{2H_\tau}/2, \tag{A13}$$

where

$$\mu_\tau(\mathbf{x},\mathbf{y}) = m_\tau(\mathbf{x})m_\tau(\mathbf{y}) + (\sigma_\tau^2(\mathbf{x})+\sigma_\tau^2(\mathbf{y}))/2 - \beta_\tau(\mathbf{x},\mathbf{y})/2. \tag{A14}$$

As a result, the key relation (17) becomes

$$A^{-2}\omega^3 E|\hat{u}(\omega)|^2 \approx \iint_{\Sigma\times\Sigma} \phi(\mathbf{x},\mathbf{y})\exp(-i\omega((\mathbf{x}-\mathbf{y}),\mathbf{d}))\sigma(d\mathbf{x})\sigma(d\mathbf{y}) \doteq I(\omega|\phi), \ \omega >> 1. \tag{A15}$$

where $\phi(\mathbf{x},\mathbf{y}) = m_\tau(\mathbf{x},\mathbf{y})/b_\xi(\mathbf{x},\mathbf{y})$ (see (A13) and (A11)).

By (C2), $\phi(\mathbf{x},\mathbf{y})$ is a smooth function out of the 'diagonal' $D: \mathbf{x}=\mathbf{y}$. Therefore the main contribution in the asymptotics of $I(\omega|\phi)$ comes from $D$ and the boundary frame of $\Sigma\times\Sigma$, $\partial\Sigma\times\partial\Sigma$.

### *A2.1 Contribution of inner points of* $\Sigma$.

We confine $\phi(\mathbf{x},\mathbf{y})$ within a small vicinity of $D$ with the closure of it lying strictly within $\Sigma\times\Sigma$. This is achieved by multiplying $\phi(\mathbf{x},\mathbf{y})$ by a finite smooth function that is concentrated in a suitable vicinity of $D$ and that is 1 in a smaller vicinity of $D$ For the sake of simplicity we retain the notation for the modified function $\phi(\mathbf{x},\mathbf{y})$ , but we shall neglect the behavior of $\phi(\mathbf{x},\mathbf{y})$ at the boundary of the vicinity of $D$. This will enable us to omit any mention of the integration domain in what follows.

By (A11),

$$b_\xi^{-1}(\mathbf{x},\mathbf{y}) = (\Delta_\xi^2(\mathbf{x},\mathbf{y})\sigma_+^2(\mathbf{x},\mathbf{y}) - \sigma_-^2(\mathbf{x},\mathbf{y}) - \Delta_\xi^4(\mathbf{x},\mathbf{y}))^{-1/2}, \tag{A16}$$

$$\sigma_+^2(\mathbf{x},\mathbf{y}) = 2(\sigma_\xi^2(\mathbf{x})+\sigma_\xi^2(\mathbf{y})) \approx 4\sigma_\xi^2(\mathbf{x}) > 0, \mathbf{x}-\mathbf{y}\to\mathbf{0}, \tag{A17}$$

$$\sigma_-^2(\mathbf{x},\mathbf{y}) = (\sigma_\xi^2(\mathbf{x})-\sigma_\xi^2(\mathbf{y}))^2 \approx (\nabla\sigma_\xi^2(\mathbf{x}),\mathbf{x}-\mathbf{y})^2, \mathbf{x}-\mathbf{y}\to\mathbf{0}. \tag{A18}$$

Substituting (A13) and (A16, A12) in $m_\tau(\mathbf{x},\mathbf{y})/b_\xi(\mathbf{x},\mathbf{y})$, one has

$$\phi(\mathbf{x},\mathbf{y}) = |\mathbf{x}-\mathbf{y}|^{-H_\xi} f(\mathbf{x},\mathbf{y}), \tag{A19}$$

$$f(\mathbf{x},\mathbf{y}) = (\tilde{\mu}_\tau(\mathbf{x},\mathbf{y}) - \tilde{\alpha}_\tau^2(\mathbf{x},\mathbf{y})|\mathbf{x}-\mathbf{y}|^{2H_\tau})\tilde{b}(\mathbf{x},\mathbf{y}), \tag{A20}$$

$$\tilde{\mu}_\tau(\mathbf{x},\mathbf{y}) = \mu_\tau(\mathbf{x},\mathbf{y})/[\alpha_\xi(\mathbf{x},\mathbf{y})\sigma_+(\mathbf{x},\mathbf{y})] \approx 0.5E\tau^2(\mathbf{x})[\alpha_\xi(\mathbf{x})\sigma_\xi(\mathbf{x})]^{-1} > 0, \mathbf{x}-\mathbf{y}\to\mathbf{0}, \tag{A21}$$

$$\breve{\alpha}_\tau(\mathbf{x},\mathbf{y}) = \alpha_\tau(\mathbf{x},\mathbf{y}) / [\alpha_\xi(\mathbf{x},\mathbf{y})\sigma_+(\mathbf{x},\mathbf{y})] \approx 0.5\alpha_\tau^2(\mathbf{x})[\alpha_\xi(\mathbf{x})\sigma_\xi(\mathbf{x})]^{-1} > 0\,,\ \mathbf{x}-\mathbf{y} \to \mathbf{0}\,, \tag{A22}$$

$$\breve{b}(\mathbf{x},\mathbf{y}) = (1 - a_+(\mathbf{x},\mathbf{y})|\mathbf{x}-\mathbf{y}|^{2H_\xi} - a_-(\mathbf{x},\mathbf{y})|\mathbf{x}-\mathbf{y}|^{-2H_\xi} - r(\mathbf{x},\mathbf{y}))^{-1/2}\,, \tag{A23}$$

where

$$a_+(\mathbf{x},\mathbf{y}) = \alpha_\xi^2(\mathbf{x},\mathbf{y}) / \sigma_+^2(\mathbf{x},\mathbf{y}) \approx \alpha_\xi^2(\mathbf{x})/(4\sigma_\xi^2(\mathbf{x})) > 0\,, \tag{A24}$$

$$a_-(\mathbf{x},\mathbf{y}) = [-\beta_\xi(\mathbf{x},\mathbf{y})\sigma_+^2(\mathbf{x},\mathbf{y}) + \sigma_-^2(\mathbf{x},\mathbf{y}) + \beta_\xi^2(\mathbf{x},\mathbf{y})]/[\alpha_\xi^2(\mathbf{x},\mathbf{y})\sigma_+^2(\mathbf{x},\mathbf{y})] = O(|\mathbf{x}-\mathbf{y}|^2)\,, \tag{A25}$$

$$r(\mathbf{x},\mathbf{y}) = 2\beta_\xi(\mathbf{x},\mathbf{y}) / \sigma_+^2(\mathbf{x},\mathbf{y}) = O(|\mathbf{x}-\mathbf{y}|^2)\,. \tag{A26}$$

Integral (A15) is

$$I(\omega|\breve{\phi}) = \int \breve{\phi}(\mathbf{z}) \exp(i\omega(\mathbf{z},\mathbf{d}))\sigma(d\mathbf{z}) \quad , \tag{A27}$$

where

$$\breve{\phi}(\mathbf{z}) = \int \phi(\mathbf{x},\mathbf{x}-\mathbf{z})\sigma(d\mathbf{x}) = |\mathbf{z}|^{-H_\xi} \breve{f}(\mathbf{z}) \ ,\ \mathbf{z} = (z', z'')\,, \tag{A28}$$

$$\breve{f}(\mathbf{z}) = \int f(\mathbf{x},\mathbf{x}-\mathbf{z})\sigma(d\mathbf{x})\,, \tag{A29}$$

and $f(\mathbf{x},\mathbf{y})$ is given by (A20). Formal application of the result (A1-A4) to $I(\omega|\breve{\phi})$ gives the desired asymptotics :

$$I(\omega|\breve{\phi}) \approx (\omega|\mathbf{d}|)^{-(2-H_\xi)} 2\pi L(-H_r)\breve{f}(0)\,, \tag{A30}$$

$$\breve{f}(0) = \int_\Sigma \breve{\mu}_\tau(\mathbf{x},\mathbf{x})\sigma(d\mathbf{x}) = 1/2 \int_\Sigma E\tau^2(\mathbf{x})[\sigma_\xi(\mathbf{x})\alpha_\xi(\mathbf{x})]^{-1}\sigma(d\mathbf{x})\,. \tag{A31}$$

To justify this approach, we represent $\phi(\mathbf{x},\mathbf{y})$ near $D$ by a finite sum $\sum \phi_\alpha(\mathbf{x},\mathbf{y})$ and verify conditions (A2, A3) for each summand separately. We show that asymptotics (A30) is dominant for the integrals $I(\omega|\phi_\alpha)$.

***Local structure of*** $\phi(\mathbf{x},\mathbf{y})$.

By (A23-A26),

$$\breve{b}(\mathbf{x},\mathbf{y}) = (1 - \breve{a}_+ - \breve{a}_- - r)^{-1/2} \quad , \tag{A32}$$

where

$$r = r(\mathbf{x},\mathbf{y}) = O(|\mathbf{x}-\mathbf{y}|^2\,, \tag{A33}$$

$$\breve{a}_+ = a_+(\mathbf{x},\mathbf{y})|\mathbf{x}-\mathbf{y}|^{2H_\xi} = O(|\mathbf{x}-\mathbf{y}|^{2H_\xi})\,, \tag{A34}$$

$$\breve{a}_- = a_-(\mathbf{x},\mathbf{y})|\mathbf{x}-\mathbf{y}|^{-2H_\xi} = O(|\mathbf{x}-\mathbf{y}|^{2\overline{H}_\xi}),\ \overline{H}_\xi = 1 - H_\xi\,. \tag{A35}$$

Since (A34, A35) is small near $D$, we can choose a neighborhood $O_D$ of $D$ such that $|\tilde{a}_+| + |\tilde{a}_-| + |r| < 1$.

Then we may represent $\tilde{b}(\mathbf{x},\mathbf{y})$ by an absolutely convergent series in $O_D$:

$$\tilde{b}(\mathbf{x},\mathbf{y}) = \sum_{n\geq 0} \rho_n (\tilde{a}_+ + \tilde{a}_- + r)^n = \sum \tilde{\rho}_{k_1k_2k_3} \tilde{a}_+^{k_1} \tilde{a}_-^{k_2} r^{k_3} , \quad \text{(A36)}$$

where

$$\tilde{\rho}_{k_1k_2k_3} = (1/2)_n / [k_1! k_2! k_3!], n = k_1 + k_2 + k_3; (a)_n = \Gamma(a+n)/\Gamma(a). \quad \text{(A37)}$$

Obviously, $\tilde{\rho}_{k,0,0} = \rho_k = (1/2)_k / k!$

Because (A36) is an absolutely convergent series, any grouping and permutation of the series terms are admissible. In addition, if $|u| < 1$ we can repeatedly differentiate both sides of the relation $\sum_{n\geq 0} \rho_n u^n = (1-u)^{-1/2}$ . Consequently, we can repeatedly differentiate (A36) with respect to $\mathbf{x},\mathbf{y} \in O_D \setminus D$ because $u = \tilde{a}_+ + \tilde{a}_- + r$ is a smooth function in this domain.

Note, that $\tilde{a}_+\tilde{a}_- = a_+a_-$ and $a_+a_-, r$ are smooth functions of the type $\varphi = O(|\mathbf{x}-\mathbf{y}|^2)$ Therefore $\tilde{a}_+^{n_1}\tilde{a}_-^{n_2} r^{n_3} = (\tilde{a}_\pm)^i \varphi, i = |n_1 - n_2|$ , where $\varphi$ is a smooth function of the type $\varphi = O(|\mathbf{x}-\mathbf{y}|^2)$ as well. As a result, one has

$$\tilde{b}(\mathbf{x},\mathbf{y}) = 1 + u(\mathbf{x},\mathbf{y}) + \sum_{j=1}^{k_+} \tilde{a}_+^j(\mathbf{x},\mathbf{y})(\rho_j + \varphi_j^+(\mathbf{x},\mathbf{y})) + \sum_{j=1}^{k_-} \tilde{a}_-^j(\mathbf{x},\mathbf{y})(\rho_j + \varphi_j^-(\mathbf{x},\mathbf{y})). \quad \text{(A38)}$$

Here $u$ joins all monomials $\tilde{a}_+^{n_1}\tilde{a}_-^{n_2} r^{n_3}$ from (A36) with $n_1 = n_2, n_1 + n_2 + n_3 > 0$. Therefore $u$ is a smooth function of the type $\varphi = O(|\mathbf{x}-\mathbf{y}|^2)$

The jth term of the 1st sum in (A38) includes all terms $\tilde{a}_+^{n_1}\tilde{a}_-^{n_2} r^{n_3}$ with $n_1 - n_2 = j > 0$, while the jth term of the 2d sum includes the similar terms with $n_1 - n_2 = -j < 0$. Therefore

$$\varphi_j^\pm(\mathbf{x},\mathbf{y}) = \sum_{l>0} (\tilde{a}_\pm(\mathbf{x},\mathbf{y}))^{lk_\pm} (\rho_{j+lk_\pm} + u_{j,l}^\pm(\mathbf{x},\mathbf{y})), \quad \text{(A39)}$$

where $u_{j,l}^\pm$ are smooth functions of the type $\varphi = O(|\mathbf{x}-\mathbf{y}|^2)$. The numbers $k_\pm$ are such that

$$1 \leq 2h_\pm k_\pm < 1 + 2h_\pm \text{ , where } h_+ = H_\xi,\ h_- = 1 - H_\xi . \quad \text{(A40)}$$

As a result,

$$\varphi_j^\pm(\mathbf{x},\mathbf{y}) = O((\tilde{a}_\pm)^{k_\pm}) = O(|\mathbf{x}-\mathbf{y}|^{2h_\pm k_\pm}) = O(|\mathbf{x}-\mathbf{y}|) \text{ , } \mathbf{x}-\mathbf{y} \to \mathbf{0}. \quad \text{(A41)}$$

Finally, (A20, A38) give us the desired representation of $\phi(\mathbf{x},\mathbf{y}) = |\mathbf{x}-\mathbf{y}|^{-H_\xi} f(\mathbf{x},\mathbf{y})$ by a finite sum of $\phi_\alpha(\mathbf{x},\mathbf{y})$.

***The spectrum asymptotics of the*** $\phi(\mathbf{x},\mathbf{y})$ ***components*** (see (A20, A38).

Assuming that $\phi = \sum \phi_\alpha(\mathbf{x},\mathbf{y})$, our goal is the asymptotics of integral (A27) for $\breve{\phi}_\alpha(\mathbf{z})$ (see (A28) for notation).

**The 1st type of** $\phi_\alpha(\mathbf{x},\mathbf{y})$.

The principal component of $\phi(\mathbf{x},\mathbf{y})$ is $\phi_0(\mathbf{x},\mathbf{y}) = |\mathbf{z}|^h f_0(\mathbf{x},\mathbf{y})$, $\mathbf{x}-\mathbf{y} = \mathbf{z}$, $h = -H_\xi$ where

$$f_0(\mathbf{x},\mathbf{y}) = (\tilde{\mu}_\tau(\mathbf{x},\mathbf{y}) - \tilde{\alpha}_\tau^2(\mathbf{x},\mathbf{y})|\mathbf{z}|^{2H_\tau}/2)(1+u(\mathbf{x},\mathbf{y})). \tag{A42}$$

***The case*** $(2H_\tau \ge 1)$. In order to use the result (A1-A4), we write $\breve{\phi}_0(\mathbf{z}) = \breve{f}_0(0)|\mathbf{z}|^h + \varphi(\mathbf{z})$, where $\varphi(\mathbf{z}) = (\breve{f}_0(\mathbf{z}) - \breve{f}_0(0))|\mathbf{z}|^h$ and

$$\breve{f}_0(0) = \int f_0(\mathbf{x},\mathbf{x})\sigma(d\mathbf{x}) = \int \tilde{\mu}_\tau(\mathbf{x},\mathbf{x})\sigma(d\mathbf{x}). \tag{A43}$$

In the notation of statement (A1-A4), we may use $q = h$ and $m = 1$, because $h \in (-1,1)$. We have to verify that $\varphi(\mathbf{z}) = O(|\mathbf{z}|^{h+1})$ and $\Delta\varphi(\mathbf{z}) = O(|\mathbf{z}|^{h-1})$. The first condition follows from the relation:

$$\begin{aligned}\breve{f}_0(\mathbf{z}) - \breve{f}_0(0) &= \int (\tilde{\mu}_\tau(\mathbf{x},\mathbf{x}-\mathbf{z}) - \tilde{\mu}_\tau(\mathbf{x},\mathbf{x}))\sigma(d\mathbf{x}) \\ &+ \int (\tilde{\mu}_\tau(\mathbf{x},\mathbf{x}-\mathbf{z})u(\mathbf{x},\mathbf{x}-\mathbf{z}))\sigma(d\mathbf{x}) - |\mathbf{z}|^{2H_\tau} \int (\tilde{\alpha}_\tau^2(\mathbf{x},\mathbf{x}-\mathbf{z})(1+u(\mathbf{x},\mathbf{x}-\mathbf{z}))\sigma(d\mathbf{x})/2 \\ &= O(|\mathbf{z}|) + O(|\mathbf{z}|^2) + O(|\mathbf{z}|^{2H_\tau}) = O(|\mathbf{z}|).\end{aligned} \tag{A44}$$

Here we use smoothness of $\tilde{\mu}_\tau(\mathbf{x},\mathbf{y})$ (see (A21), relations $u(\mathbf{x},\mathbf{y}) = O(|\mathbf{x}-\mathbf{y}|^2)$ and $2H_\tau \ge 1$.

Function $\varphi(\mathbf{z}) = (\breve{f}_0(\mathbf{z}) - \breve{f}_0(0))|\mathbf{z}|^h$ is smooth out of $\mathbf{z} = 0$ and

$$\Delta\varphi(\mathbf{z}) = \Delta[|\mathbf{z}|^h(\breve{f}_0(\mathbf{z}) - \breve{f}_0(0))] = O(\Delta|\mathbf{z}|^h)O(\breve{f}_0(\mathbf{z}) - \breve{f}_0(0)) + |\mathbf{z}|^h O(\Delta\breve{f}_0(z)). \tag{A45}$$

Applying the Laplace operator to both sides of (A44), we have $O(\Delta\breve{f}_0(\mathbf{z})) = O(1) + O(1) + O(\Delta|\mathbf{z}|^{2H_\tau})$.

But $\Delta|\mathbf{z}|^h = h^2|\mathbf{z}|^{h-2}$ and $2H_\tau \ge 1$. Hence $O(\Delta\breve{f}_0(\mathbf{z})) = O(|\mathbf{z}|^{2H_\tau-2}) = O(|\mathbf{z}|^{-1})$ and $\Delta\varphi(\mathbf{z}) = O(|\mathbf{z}|^{h-1})$.

Due to (A1-A4), we get the asymptotics (A30) for the component $\phi_0(\mathbf{x},\mathbf{y})$ of $\phi(\mathbf{x},\mathbf{y})$. More precisely,

$$I(\omega|\phi_0) = (\omega|\mathbf{d}|)^{-(2-H_\xi)} 2\pi L(-H_\xi)\breve{f}(0) + o(|\omega\mathbf{d}|^{-2}), \tag{A46}$$

where $\breve{f}(0)$ is given by (A31).

***The case*** ($2H_\tau < 1$). One has $\phi_0(\mathbf{x},\mathbf{y}) = |\mathbf{z}|^{h_1} f_1(\mathbf{x},\mathbf{y}) + |\mathbf{z}|^{h_2} f_2(\mathbf{x},\mathbf{y})$, where

$$f_1(\mathbf{x},\mathbf{y}) = \tilde{\mu}_\tau(\mathbf{x},\mathbf{y})(1+u(\mathbf{x},\mathbf{y})), h_1 = -H_\xi \in (-1,1), \tag{A47}$$

$$f_2(\mathbf{x},\mathbf{y}) = -\tilde{\alpha}_\tau^2(\mathbf{x},\mathbf{y})(1+u(\mathbf{x},\mathbf{y})/2), h_2 = -H_\xi + 2H_\tau \in (-1,1). \tag{A48}$$

The preceding argument is still valid for the above terms of $\phi_0(\mathbf{x},\mathbf{y})$ with the parameters $q = h = h_\alpha, \alpha = 1,2$, and $m = 1$. Therefore

$$I(\omega|\phi_0) = \sum_\alpha (\omega|\mathbf{d}|)^{-2-h_\alpha} 2\pi L(h_\alpha)\breve{f}_\alpha(0) + o(|\omega\mathbf{d}|^{-2}). \tag{A49}$$

Note, that $h_1 < \min(h_2,2)$ and $\breve{f}_1(0) = \breve{f}(0)$. Therefore we can continue

$$I(\omega|\phi_0) = (\omega|\mathbf{d}|)^{-2+H_\xi} 2\pi L(-H_\xi)\breve{f}(0)(1+O(|\omega\mathbf{d}|^{-\min(H_\xi,2H_\tau)})). \tag{A50}$$

**The 2d type of $\phi_\alpha(\mathbf{x},\mathbf{y})$.**

The summands of the 1st sum in (A31) generate the following components of $\phi(\mathbf{x},\mathbf{y})$:

$$\phi_j^+(\mathbf{x},\mathbf{y}) = |z|^{-H_\xi} (\tilde{\mu}_\tau(\mathbf{x},\mathbf{y}) - \tilde{\alpha}_\tau^2(\mathbf{x},\mathbf{y})|\mathbf{z}|^{2H_\tau}/2)(\rho_j + \varphi_j^+(\mathbf{x},\mathbf{y}))\tilde{a}_+^j(\mathbf{x},\mathbf{y}) = |\mathbf{z}|^{h_i} f_+(\mathbf{x},\mathbf{y}). \tag{A51}$$

Here

$$f_+(\mathbf{x},\mathbf{y}) = (\tilde{\mu}_\tau(\mathbf{x},\mathbf{y}) - \tilde{\alpha}_\tau^2(\mathbf{x},\mathbf{y})|\mathbf{z}|^{2H_\tau}/2)(\rho_j + \varphi_j^+(\mathbf{x},\mathbf{y}))a_+^j(\mathbf{x},\mathbf{y}), \tag{A52}$$

$\tilde{\mu}_\tau, \tilde{\alpha}_\tau^2, a_+$ are smooth functions, $\varphi_j^+$ is given by (A39).

Due to (A34), $h_j = (2j-1)H_\xi, j = 1,...k_+$. By definition of $k_+$, (A 40), one has

$$h_j \in (0,1), j < k_+ \text{ and } |\ h_{k_+} - 1\ | < H_\xi. \tag{A53}$$

***The case*** ($2H_\tau \ge 1$). In order to use the result (A1-A 4) we set $q = h = h_j$, $m = m(j)$, where

$$m(j) = 1 \text{ for } h \in (-1,1) \text{ (It is the case } j < k_+\text{); and } m(j) = 2 \text{ for } h \in [1,3). \tag{A54}$$

The last case is possible if $h_{k_+} \ge 1$. As above (see section "The 1st type of $\phi_\alpha(\mathbf{x},\mathbf{y})$", the case $2H_\tau \ge 1$), we write $\breve{\phi}_j^+(\mathbf{z}) = \breve{f}_+(0)|\mathbf{z}|^h + \varphi(\mathbf{z})$, where $\varphi(\mathbf{z}) = (\breve{f}_+(z) - \breve{f}_+(0))|\mathbf{z}|^h$. One has

$$\breve{f}_+(\mathbf{z}) - \breve{f}_+(0) = \rho_j \int (\tilde{\mu}_\tau(\mathbf{x},\mathbf{x}-\mathbf{z})a_+^j(\mathbf{x},\mathbf{x}-\mathbf{z}) - \tilde{\mu}_\tau(\mathbf{x},\mathbf{x})a_+^j(\mathbf{x},\mathbf{x}))\sigma(d\mathbf{x})$$

$$+ \int \tilde{\mu}_\tau(\mathbf{x},\mathbf{x}-\mathbf{z})a_+^j(\mathbf{x},\mathbf{x}-\mathbf{z})\varphi_j^+(\mathbf{x},\mathbf{x}-\mathbf{z})\sigma(d\mathbf{x})$$

$$-|\mathbf{z}|^{2H_\tau}\int(\widetilde{\alpha}_\tau^2(\mathbf{x},\mathbf{x}-\mathbf{z})a_+^j(\mathbf{x},\mathbf{x}-\mathbf{z})(\rho_j+\varphi_j^+(\mathbf{x},\mathbf{x}-\mathbf{z}))\sigma(d\mathbf{x})/2$$

$$=O(|\mathbf{z}|)+O(|\mathbf{z}|)+O(|\mathbf{z}|^{2H_\tau})=O(|\mathbf{z}|) \tag{A55}$$

Here we use smoothness of $\widetilde{\mu}_\tau(\mathbf{x},\mathbf{y})$ and $a_+(\mathbf{x},\mathbf{y})$ (see (A21, A34)), the relations $\varphi_i^+(\mathbf{x},\mathbf{y})=O(|\mathbf{x}-\mathbf{y}|\ )$ (see (A41)), and $2H_\tau \geq 1$. Due to (A55), $\varphi(\mathbf{z})=(\breve{f}(\mathbf{z})-\breve{f}(0))|\mathbf{z}|^h\ =O(|\mathbf{z}|^{h+1})$.

To show that $\Delta^p\varphi(\mathbf{z})=O(|\mathbf{z}|^{h-2p+1}),1\leq p\leq m$, note that

$$\varphi_j^\pm(\mathbf{x},\mathbf{x}-\mathbf{z})=\sum\nolimits_{l>0}(a_\pm(\mathbf{x},\mathbf{x}-\mathbf{z}))^{lk_\pm}(\rho_{j+lk_\pm}+u_{j,l}^\pm(\mathbf{x},\mathbf{x}-\mathbf{z}))|\mathbf{z}|^{2h_\pm lk_\pm}, \tag{A56}$$

where $a_\pm, u_{j,l}^\pm$ are smooth functions. Therefore

$$\Delta^p\varphi_i^\pm(\cdot,\cdot-\mathbf{z})=O(|\mathbf{z}|^{2h_\pm k_\pm-2p})=O(|\mathbf{z}|^{1-2p}). \tag{A57}$$

Due to (A55) and (A57), one has $O(\Delta^p\breve{f}_+(\mathbf{z}))=O(1)+O(|\mathbf{z}|^{1-2p})+O(|\mathbf{z}|^{2H_\tau-2p})=O(|\mathbf{z}|^{1-2p})$ Therefore

$$\Delta^p\varphi(\mathbf{z})=O(\Delta^p|\mathbf{z}|^h)O(\breve{f}_+(\mathbf{z})-\breve{f}_+(0))+O(\Delta^p\breve{f}_+(\mathbf{z}))|\mathbf{z}|^h=O(|\mathbf{z}|^{h-2p+1}),1\leq p\leq m. \tag{A58}$$

The result is

$$I(\omega|\phi_j^+)=(\omega|\mathbf{d}|)^{-2-(2j-1)H_\xi}\,2\pi L((2j-1)H_\xi)\breve{f}_+(0)+o(|\omega\mathbf{d}|^{-2m(j)}) \tag{A59}$$

(see (A54) for $m(j)$). The first term of this asymptotics becomes nontrivial only if $m(j)=2$ only, because $2+(2i-1)H_\xi>2m(j)$ if $m(j)=1$. On the whole $I(\omega|\phi_i^+)=o(|\omega\mathbf{d}|^{-2})$.

***The case*** $(2H_\tau<1)$. As above, we split $\phi_j^+(\mathbf{x},\mathbf{y})$ into two parts

$$\phi_j^+(\mathbf{x},\mathbf{y})=|z|^{-H_\xi}(\widetilde{\mu}_\tau(\mathbf{x},\mathbf{y})(\rho_i+\varphi_j^+(\mathbf{x},\mathbf{y}))\widetilde{a}_+^j(\mathbf{x},\mathbf{y})-|\mathbf{z}|^{2H_\tau-H_\xi}\widetilde{\alpha}_\tau^2(\mathbf{x},\mathbf{y})(\rho_j+\varphi_j^+(\mathbf{x},\mathbf{y}))\widetilde{a}_+^j(\mathbf{x},\mathbf{y})/2 \tag{A60}$$

and find the spectrum asymptotics for each part separately using previous arguments. As a result, we come to the relation

$$I(\omega|\phi_j^+)=o(|\omega\mathbf{d}|^{-2}). \tag{A61}$$

**The 3d type of $\phi_\alpha(\mathbf{x},\mathbf{y})$.**

The summands of the 2d sum in (A31) generate the following type of $\phi_\alpha(\mathbf{x},\mathbf{y})$:

$$\phi_j^-(\mathbf{x},\mathbf{y})=|z|^{-H_\xi}(\widetilde{\mu}_\tau(\mathbf{x},\mathbf{y})-\widetilde{\alpha}_\tau^2(\mathbf{x},\mathbf{y})|\mathbf{z}|^{2H_\tau}/2)(\rho_j+\varphi_j^-(\mathbf{x},\mathbf{y}))\widetilde{a}_-^j(x,y)=|\mathbf{z}|^{h_i}f_-(\mathbf{x},\mathbf{y})a_-^j(\mathbf{x},\mathbf{y}) \tag{A62}$$

where $h_j=-(2j+1)H_\xi$,

$$f_-(\mathbf{x},\mathbf{y}) = (\widetilde{\mu}_\tau(\mathbf{x},\mathbf{y}) - \widetilde{\alpha}_\tau^2(\mathbf{x},\mathbf{y})|\mathbf{z}|^{2H_\tau}/2)(\rho_j + \varphi_j^-(\mathbf{x},\mathbf{y})), \tag{A63}$$

$$a_-(\mathbf{x},\mathbf{y}) = [-\beta_\xi(\mathbf{x},\mathbf{y})\sigma_+^2(\mathbf{x},\mathbf{y}) + \sigma_-^2(\mathbf{x},\mathbf{y}) + \beta_\xi^2(\mathbf{x},\mathbf{y})]/[\alpha_\xi^2(\mathbf{x},\mathbf{y})\sigma_+^2(\mathbf{x},\mathbf{y})] = O(|\mathbf{x}-\mathbf{y}|^2). \tag{A64}$$

Therefore, $\phi_i^-(\mathbf{x},\mathbf{y}) = O(|\mathbf{z}|^{h_i+2j}) = O(|\mathbf{z}|^{2\overline{H}_\xi j - H_\xi})$, $j = 1,...k_-$, $\overline{H}_\xi = 1 - H_\xi$, where $2j\overline{H}_\xi - H_\xi \in (-1,3)$, see (A40). By conditions (C2-C4),

$$a_-(\mathbf{x},\mathbf{y}) = (W(\mathbf{x},\mathbf{y})\mathbf{z},\mathbf{z}) + V(\mathbf{x},\mathbf{y}), \mathbf{z} = \mathbf{x} - \mathbf{y}, \tag{A65}$$

where $V(\mathbf{x},\mathbf{y})$ and the elements of $2\times 2$ matrix $W(\mathbf{x},\mathbf{y})$ are smooth functions. In addition, $V(\mathbf{x},\mathbf{y}) = O(|\mathbf{x}-\mathbf{y})|^3)$.

Because of (A62), (A65), we write

$$\breve{\phi}_j^-(\mathbf{z}) = |\mathbf{z}|^{h_i} \sum\nolimits_{n_1+n_2=2j} w_{n_1 n_2} z_1^{n_1} z_2^{n_2} + \varphi(\mathbf{z}), \quad \mathbf{z} = (z_1, z_2), \tag{A66}$$

where

$$\sum\nolimits_{n_1+n_2=2j} w_{n_1 n_2} z_1^{n_1} z_2^{n_2} = \rho_i \int \widetilde{\mu}_\tau(\mathbf{x},\mathbf{x})[(W(\mathbf{x},\mathbf{x})\mathbf{z},\mathbf{z})]^j \sigma(d\mathbf{x}) \tag{A67}$$

and $\varphi(\mathbf{z}) = O(|\mathbf{z}|^{h_i+2j+1})$.

For simplicity we suppose that $w_{n_1 n_2} \neq 0$ for some $(n_1, n_2)$ .

In the case $2H_\tau \geq 1$ we choose the following parameters:

$$h = h_j, q = h_j + 2j = 2j\overline{H}_\xi - H_\xi \in (-1,3); m = m(j),$$

where $m(j) = 1$ if $q \in (-1,1)$ and $m(j) = 2$ if $q \in [1,3)$.

It remains to verify the conditions: $\Delta^p \varphi(\mathbf{z}) = O(|z|^{q-2p+1}),\ p = 0,1,...m(j)$.

This can be done with the help of (A57) in the same manner as above.

In the case $2H_\tau < 1$ we split (A63) into two part and consider separately two summands of $\phi_j^-(\mathbf{x},\mathbf{y})$, namely

$$\phi_{j,\alpha}^-(\mathbf{x},\mathbf{y}) = |\mathbf{z}|^{h_j(\alpha)} f_\alpha(\mathbf{x},\mathbf{y}) a_-^j(\mathbf{x},\mathbf{y}), \alpha = 1,2, \tag{A68}$$

where

$$f_1(\mathbf{x},\mathbf{y}) = (\widetilde{\mu}_\tau(\mathbf{x},\mathbf{y})(\rho_j + \varphi_j^-(\mathbf{x},\mathbf{y})),\ \ h_j(1) = -(2j+1)H_\xi \tag{A69}$$

$$f_2(\mathbf{x},\mathbf{y}) = -0.5\widetilde{\alpha}_\tau^2(\mathbf{x},\mathbf{y})(\rho_j + \varphi_j^-(\mathbf{x},\mathbf{y})),\ \ h_j(2) = h_j(1) + 2H_\tau, \ . \tag{A70}$$

The function $\breve{\phi}_{j,\alpha}^-(z)$ is represented by (A66, A67) if $\alpha = 1$ and by (A66, A71) if $\alpha = 2$ :

$$\sum_{n_1+n_2=2j} w_{n_1n_2} z_1^{n_1} z_2^{n_2} = -0.5\rho_j \int \tilde{\mu}_\tau(\mathbf{x},\mathbf{x})\tilde{\alpha}_\tau^2(\mathbf{x},\mathbf{x})[(W(\mathbf{x},\mathbf{x})\mathbf{z},\mathbf{z})]^j \sigma(d\mathbf{x}). \qquad \text{(A71)}$$

In the last case ($\alpha = 2$) we use $q = 2j + h_i(\alpha) \in (-1,5)$. Therefore, $m = 1,2$ or $3$ if $h_j(2) \in (-1,1),[1,3)$ or $[3,5)$ respectively.

Finely, we use (A1-A4) to derive the following asymptotics for different variants of $\breve{\phi}_j^-(z)$:

$$I(\omega|\phi_j^-) = 2\pi L(h_j)(-1)^j \sum_{n_1+n_2=2j} w_{n_1n_2} (\partial/\partial k_1)^{n_1} (\partial/\partial k_2)^{n_2} |\mathbf{k}|^{-2-h_i}|_{\mathbf{k}=\omega\mathbf{d}} + o(|\omega\mathbf{d}|^{-2m(j)})$$

$$= O(|\omega\mathbf{d}|^{-2j-h_j-2}) + o(|\omega\mathbf{d}|^2) \;. \qquad \text{(A72)}$$

But $\min(2j + h_j + 2) = \min(2j\overline{H}_\xi - H_\xi + 2) = 4 - 3H_\xi > 2 - H_\xi$. Hence

$$I(\omega|\phi_j^-) = O(|\omega d|^{-\min(4-3H_\xi,2)}) \;. \qquad \text{(A73)}$$

By bringing together (A15), (A50), (A61), (A73), we get the following result, related to the inner points of $\Sigma$ :

$$A^{-2}\omega^3 E|\hat{u}_\Sigma(\omega)|^2 = (\omega|\mathbf{d}|)^{-2+H_\xi} 2\pi L(-H_\xi)\breve{f}(0)(1 + O(|\omega d|^{-\min(H_\xi,2H_\tau)}), \qquad \text{(A74)}$$

where $\breve{f}(0)$ is given by (A31).

***A2.2 Contribution of the boundary points of $\Sigma$.***

Suppose that $\mathbf{x}_0, \mathbf{y}_0 \in \partial\Sigma$. The contribution of $(\mathbf{x}_0, \mathbf{y}_0)$ in the asymptotics of $I(\omega|\phi)$ is given by the asymptotics of $I(\omega|\phi_\bullet)$, where $\phi_\bullet$ is the restriction of $\phi$ on a small vicinity of $(\mathbf{x}_0, \mathbf{y}_0)$, i.e., $\phi_\bullet = \phi\varphi(\mathbf{x} - \mathbf{x}_0)\varphi(\mathbf{y} - \mathbf{y}_0)$, where $\varphi$ is a smooth finite function, $\varphi = 1$ for $|\mathbf{x}| < \varepsilon$ и $\varphi = 0$ for $|\mathbf{x}| > 2\varepsilon$ .

***The case* $\mathbf{x}_0 \neq \mathbf{y}_0$.**

Suppose that $\mathbf{x}_0 \neq \mathbf{y}_0$ and $\varepsilon$ is sufficiently small. Then $\phi_\bullet$ is smooth function on $\Sigma \times \Sigma$, because $|\mathbf{x} - \mathbf{y}|$ is smooth out of the diagonal $D$.

For simplisity, we start from $\phi_\bullet = a(\mathbf{x})b(\mathbf{y})$, where $a(\mathbf{x}), b(\mathbf{y})$ are smooth functions. Then

$$I(\omega|\phi_\bullet) = J(\omega|a)J(-\omega|b) \quad \text{and} \quad J(\omega|f) = \int_\Sigma f(\mathbf{x})e^{-i\omega(\mathbf{x},\mathbf{d})}\sigma(d\mathbf{x}). \qquad \text{(A75)}$$

Suppose that $\partial\Sigma$ is smooth, the points $\mathbf{x}_\alpha(\mathbf{d}) \in \partial\Sigma$ at which $\mathbf{d}$ is orthogonal to $\partial\Sigma$ are isolated, and the curvature of the boundary at $\mathbf{x}_\alpha(\mathbf{d})$ is nonzero, $\kappa_\alpha \neq 0$. Due to **2.1, 2.3**, we have the following asymptotics: $J(\omega|f) = |\omega d|^{-3/2} \sum f(\mathbf{x}_\alpha(\mathbf{d}))|2\pi/\kappa_\alpha|^{1/2} \varepsilon_\alpha$, where $\varepsilon_\alpha$ depends on $\mathbf{x}_\alpha(\mathbf{d})$ and $|\varepsilon_\alpha| = 1$. Therefore, if the vicinity of $(\mathbf{x}_0, \mathbf{y}_0)$ contains a single point $(\mathbf{x}_\alpha(\mathbf{d}), \mathbf{x}_\beta(\mathbf{d}))$, then

$$I(\omega|\phi_\bullet) \approx |\omega \mathbf{d}|^{-3} 2\pi\phi_\bullet(\mathbf{x}_\alpha, \mathbf{x}_\beta)|\kappa_\alpha \kappa_\beta|^{-1/2} \varepsilon_\alpha \bar{\varepsilon}_\beta \quad . \tag{A76}$$

Otherwise $I(\omega|\phi_\bullet) \approx O(|\omega d|^{-N+1})$, where N is the degree of smoothness of $f$.

Relation (A76) obviously holds for any finite sum of terms like $a(\mathbf{x})b(\mathbf{y})$. With the help of such sums we can approximate any smooth function together with its derivatives of finite order. Therefore (A76) holds for any smooth function $\phi_\bullet$. Covering $\partial\Sigma \times \partial\Sigma$ by small domains that contain at most one point of the type $(\mathbf{x}_\alpha(\mathbf{d}), \mathbf{x}_\beta(\mathbf{d}))$, we can find the total contribution of set $\partial\Sigma \times \partial\Sigma \setminus D$ in the asymptotics of $I(\omega|\phi)$. Namely,

$$I(\omega|\phi_{\partial\Sigma}^{co}) \approx |\omega \mathbf{d}|^{-3} \sum_{\alpha \neq \beta} 2\pi\phi(\mathbf{x}_\alpha, \mathbf{x}_\beta)|\kappa_{x_\alpha} \kappa_{x_\beta}|^{-1/2} \varepsilon_{x_\alpha} \bar{\varepsilon}_{x_\beta} \quad , \tag{A77}$$

where $\phi(\mathbf{x}, \mathbf{y}) = m_\tau(\mathbf{x}, \mathbf{y}) / b_\xi(\mathbf{x}, \mathbf{y})$. If $\partial\Sigma$ is strictly convex, the set $\{\mathbf{x}_\alpha(\mathbf{d})\}$ consists of two points: $\mathbf{x}_+, \mathbf{x}_-$. In this case (A77) can be simplified. By (Fedoryuk (1987) (Ch.3, Th. 4.3), $\varepsilon_{\mathbf{x}_\alpha} = e^{i(\mathbf{x}_\alpha, \mathbf{d}) - is_\alpha \pi/4}$, where $s_\alpha = \mathrm{sgn}(\mathbf{d}, \mathbf{n}(\mathbf{x}_\alpha))$ and $\mathbf{n}(\mathbf{x}_\alpha)$ is the outer normal to $\partial\Sigma$ at $\mathbf{x}_\alpha$. Therefore one has

$$I(\omega|\phi_{\partial\Sigma}^{co}) \approx s|\omega \mathbf{d}|^{-3} 2\pi\phi(\mathbf{x}_+, \mathbf{x}_-)|\kappa_+ \kappa_-|^{-1/2} \sin(\omega(x_+ - x_-, d)), \tag{A78}$$

$$\phi(\mathbf{x}, \mathbf{y}) = 0.5E\tau(\mathbf{x})\tau(\mathbf{y}) / (E\xi^2(\mathbf{x})E\xi^2(\mathbf{y}) - [E\xi(x)\xi(y)]^2)^{1/2} \; . \tag{A79}$$

As we can see from (A78), the right part of (A77) is alternating in sign instead of being positive. The reason is that we sum the contributions of pairs $\mathbf{x}_0, \mathbf{y}_0 \in \partial\Sigma$ with $\mathbf{x}_0 \neq \mathbf{y}_0$ only. Therefore, we automatically come to the conclusion that the contribution of $(\mathbf{x}_0, \mathbf{x}_0), \mathbf{x}_0 \in \partial\Sigma$ in the integral $I(\omega|\phi)$ has to be of the order $O(\omega^{-\theta}), \theta \leq 3$.

Now we show that $\theta \geq 2 - H_\xi$. Moreover, the contribution of the pair $(\mathbf{x}_0, \mathbf{x}_0), \mathbf{x}_0 \in \partial\Sigma$ in the integral $I(\omega|\phi)$ is substantially smaller relative to the contribution of the inner points of $\Sigma$. Therefore the asymptotics (A74) is dominant.

***The case*** $\mathbf{x}_0 = \mathbf{y}_0$.

***Step1***.Suppose that

$$I(\omega) = \iint_{\Sigma \times \Sigma} |x - y|^{-\theta} a_1(\mathbf{x}) a_2(\mathbf{y}) \exp(-i\omega((\mathbf{x} - \mathbf{y}), \mathbf{d})) \sigma(d\mathbf{x}) \sigma(d\mathbf{y}) \tag{A80}$$

and the Fourier transforms $\hat{a}_i(\mathbf{k}) = \int_\Sigma e^{-i(\mathbf{x},\mathbf{k})} a_i(\mathbf{x})\sigma(d\mathbf{x})$ are such that

$$|\hat{a}_i(\mathbf{k})| < c|\mathbf{k}|^{-\gamma_i}, \mathbf{k} \to \infty \quad . \tag{A81}$$

If $0 < \theta < 2 \le \gamma_1 + \gamma_2$ and $\|a_i\|^2 = \int_\Sigma |a_i(\mathbf{x})|^2 \sigma(d\mathbf{x}) < \infty$, then

$$|I(\omega)| < c_\theta |\omega\mathbf{d}|^{-2+\theta} \|a_1\| \cdot \|a_2\|, \quad \omega \to \infty . \tag{A82}$$

***Proof.*** It is well known, that the Fourier transform of $|\mathbf{x}|^{-\theta}$ is $2\pi L(-\theta)|\mathbf{k}|^{-2+\theta}$ (Gel'fand and Shilov, 1964). Hence

$$I(\omega) = (2\pi)^{-1} L(-\theta) \int |\mathbf{k} - \omega\mathbf{d}|^{-2+\theta} \hat{a}_1(-\mathbf{k})\hat{a}_2(\mathbf{k})\sigma(d\mathbf{k}) \quad . \tag{A83}$$

Let us split the integration space into two domains: $O_1 : |\mathbf{k}| < 0.5|\omega\mathbf{d}|$ and $O_2 : |\mathbf{k}| \ge 0.5|\omega\mathbf{d}|$. We get $I(\omega) = I_1 + I_2$ respectively. Because $|\mathbf{k} - \omega\mathbf{d}| > 0.5|\omega\mathbf{d}|$ for $\mathbf{k} \in O_1$, one has

$$I_1(\omega) < (2\pi)^{-1} L(-\theta) |0.5\omega\mathbf{d}|^{-2+\theta} \int |\hat{a}_1(\mathbf{k})\hat{a}_2(\mathbf{k})| \sigma(d\mathbf{k}) < c_1(\theta) |\omega\mathbf{d}|^{-2+\theta} \|a_1\| \|a_2\| . \tag{A84}$$

By the change of variables $\mathbf{k} = \mathbf{u}|\omega\mathbf{d}|$ in $I_2$, one has

$$|I_2| < c(\theta) |\omega\mathbf{d}|^{\theta} \int_{|u|>0.5} |\hat{a}_1(\mathbf{u}|\omega\mathbf{d}|)\hat{a}_2(\mathbf{u}|\omega\mathbf{d}|)| \|\mathbf{u} - \mathbf{d}/|\mathbf{d}|\|^{-2+\theta} \sigma(d\mathbf{u}) . \tag{A85}$$

Using the estimations (A81), we can continue

$$|I_2| < \tilde{c}(\theta) |\omega\mathbf{d}|^{\theta-\gamma_1-\gamma_2} \int_{|u|>0.5} |\mathbf{u}|^{-\gamma_1-\gamma_2} |\mathbf{u} - \mathbf{d}/|\mathbf{d}||^{-2+\theta} \sigma(d\mathbf{u}) \quad . \tag{A86}$$

The right part of this relation is finite if $\gamma_1 + \gamma_2 > \theta > 0$. In addition, the relation $|I_2| = O(|\omega\mathbf{d}|^{-2+\theta})$ holds if $\gamma_1 + \gamma_2 > 2$. This proves (A82).

***Step* 2**. The conditions of the previous statement are satisfied, if the boundary $\partial\Sigma$ is smooth, $\{a_i(\mathbf{x})\}$ are continues on $\Sigma$, and $\int_\Sigma |\nabla a_i(\mathbf{x})| \sigma(d\mathbf{x}) < \infty$.

Indeed, $\hat{a}_i(\mathbf{k}) = i|\mathbf{k}|^{-2} \int_\Sigma (\mathbf{k}, \nabla_\mathbf{x}) e^{-i(\mathbf{x},\mathbf{k})} a_i(\mathbf{x})\sigma(d\mathbf{x})$. Integrating by parts, one has

$$\hat{a}_i(\mathbf{k}) = -i|\mathbf{k}|^{-1} [\int_\Sigma (\mathbf{k}/|\mathbf{k}|, \nabla_\mathbf{x} a_i(\mathbf{x})) e^{-i(\mathbf{x},\mathbf{k})} \sigma(d\mathbf{x}) + \int e^{-i(\mathbf{x},\mathbf{k})} a(\mathbf{x})(\mathbf{k}/|\mathbf{k}|, \nabla P(\mathbf{x}))\delta(P(\mathbf{x}))\sigma(d\mathbf{x}), \tag{A87}$$

where $P(\mathbf{x})$ is a smooth function near $\partial\Sigma$ such that $P(\mathbf{x}) = 0$ is an equation of $\partial\Sigma$ and $\delta(P(\mathbf{x}))$ is the $\delta$-function of $\partial\Sigma$ (Gel'fand and Shilov, 1964). For example, if $x_2 = p(x_1)$ is the local equation of $\partial\Sigma$, then $\nabla P(\mathbf{x}) = (p'(x_1),-1)$ and $\delta(P(\mathbf{x}))\sigma(d\mathbf{x}) = -dx_1$. Relation (A87) proves the second statement because one has $\hat{a}_1(\mathbf{k})\hat{a}_2(-\mathbf{k}) = O(|\mathbf{k}|^{-2})$ .

**Step 3**. The previous conclusions hold if we replace $a_1(\mathbf{x})a_2(\mathbf{y})$ in (A80) by any function $f(\mathbf{x,y})$ such that $f(\mathbf{x,y})$ is continuous, $(\partial^2 / \partial x_i \partial y_j) f(\mathbf{x,y}), \mathbf{x} \neq \mathbf{y}$ are absolutely integrable in $\Sigma \times \Sigma$, and $|\nabla_{\mathbf{x}} f|, |\nabla_{\mathbf{y}} f|$ are integrable in $\Sigma \times \partial\Sigma$ and $\partial\Sigma \times \Sigma$ respectively. Because $\Delta|\mathbf{x}|^h = c|\mathbf{x}|^{h-2}$ is locally integrable function, these conditions hold for $f(\mathbf{x,y}) = |\mathbf{x}-\mathbf{y}|^h \psi(\mathbf{x,y}), h > 0$ , where $\psi(\mathbf{x,y})$ is a smooth function.

**Step 4**.Consider the integral $I(\omega|\phi_\bullet)$, where $\phi_\bullet$ is the restriction of $\phi$ on a small vicinity of $(\mathbf{x}_0,\mathbf{x}_0)$, $\mathbf{x}_0 \in \partial\Sigma$. Due to (A20, A38), $\phi_\bullet = |\mathbf{x}-\mathbf{y}|^{-H_\xi} f(\mathbf{x,y})$ , where $f(\mathbf{x,y})$ is approximated by sum $\sum |\mathbf{x}-\mathbf{y}|^{h_i} \psi_j(\mathbf{x,y})$ , $h_j > 0$ , $\psi_j$ are smooth functions. Hence, $|I(\omega|\phi_\bullet)| < c|\omega\mathbf{d}|^{-2+H_\xi} \|f\|$ where **the** norm is related to continuous function in $\varepsilon$ -vicinity of $(\mathbf{x}_0,\mathbf{x}_0) \in \Sigma \times \Sigma$ , i.e., $\|f\|$ is small. It means that the asymptotics (A74) is dominant.

## A3. The linear front, $\delta t_r(\mathbf{x}) = 0$.

In this case

$$A^{-2}\omega^2 E|\hat{u}(\omega)|^2 = \iint_{\Sigma\times\Sigma} \phi(\mathbf{x,y}) \exp(-i\omega((\mathbf{x}-\mathbf{y}),\mathbf{d}))\sigma(d\mathbf{x})\sigma(d\mathbf{y}) \doteq I(\omega|\phi), \tag{A88}$$

where

$$\phi = m_\tau(\mathbf{x,y}) = \mu_\tau(\mathbf{x,y}) - 0.5\alpha_\tau^2(\mathbf{x,y})|\mathbf{x}-\mathbf{y}|^{2H_\tau}, \tag{A89}$$

and $\mu_\tau(\mathbf{x,y}), \alpha_\tau^2(\mathbf{x,y})$ are smooth functions (see (17), (A13)). Integral (A88) is identical with (A15) when $b_\xi(x,y) = 1$ and $H_\xi = 0$. Therefore, we can use the results from Appendix 2.

***Contribution of inner points of*** $\Sigma$.

Because $\mu_\tau(\mathbf{x,y})$ is smooth, it is enough to consider the second summand in (A89) only. Let $\psi_\varepsilon(\mathbf{x})$ be a smooth function such that $\psi_\varepsilon(\mathbf{x}) = 0$ in the $\varepsilon$-neighborhood of $\partial\Sigma$ and $\psi_\varepsilon(\mathbf{x}) = 1$ if the $2\varepsilon$-neighborhood of $\mathbf{x}$ belong to $\Sigma$, where $\varepsilon$ is small. We consider $f_\varepsilon = -0.5\alpha_\tau^2(\mathbf{x,y})\psi_\varepsilon(\mathbf{x})\psi_\varepsilon(\mathbf{y})$. The contribution of the inner points of $\Sigma$ in $I(\omega)$ is given by the relation:

$$I_{\dot{\Sigma}}(\omega) = \int |\mathbf{z}|^{2H_\tau} \breve{f}_\varepsilon(\mathbf{z}) e^{-i\omega(\mathbf{z,d})}\sigma(d\mathbf{z}) , \tag{A90}$$

where $\breve{f}_\varepsilon(\mathbf{z}) = \int f_\varepsilon(\mathbf{x}, \mathbf{x}-\mathbf{z})\sigma(d\mathbf{x})$ and

$$\breve{f}_\varepsilon(\mathbf{0}) \approx -0.5\int_\Sigma \alpha_\tau^2(\mathbf{x},\mathbf{x})\sigma(d\mathbf{x}) \ . \tag{A91}$$

Note, that $\breve{f}_\varepsilon(\mathbf{z}) = \breve{f}_\varepsilon(-\mathbf{z})$ because $f_\varepsilon(\mathbf{x},\mathbf{y}) = f_\varepsilon(\mathbf{y},\mathbf{x})$. Indeed,

$$\breve{f}_\varepsilon(\mathbf{z}) = \int f_\varepsilon(\mathbf{x},\mathbf{x}-\mathbf{z})\sigma(d\mathbf{x}) = \int f_\varepsilon(\mathbf{y}+\mathbf{z},\mathbf{y})\sigma(d\mathbf{y}) = \int f_\varepsilon(\mathbf{y},\mathbf{y}+\mathbf{z})\sigma(d\mathbf{y}) = \breve{f}_\varepsilon(-\mathbf{z}). \tag{A92}$$

Therefore $\nabla\breve{f}_\varepsilon(\mathbf{0}) = -\nabla\breve{f}_\varepsilon(\mathbf{0})$, i.e. $\nabla\breve{f}_\varepsilon(\mathbf{0}) = \mathbf{0}$, because of smoothness of $\breve{f}_\varepsilon(\mathbf{z})$.

In order to use the result (A1-A4), we write

$$\phi(\mathbf{z}) = \breve{f}_\varepsilon(\mathbf{z})|\mathbf{z}|^{2H_\tau} = |\mathbf{z}|^{2H_\tau}\breve{f}_\varepsilon(\mathbf{0}) + \varphi(\mathbf{z}) \tag{A93}$$

and choose the parameters as follows: $h = 2H_\tau, q = h+1, m = 2$. Three conditions now remain to be verified: $\Delta^p\varphi(\mathbf{z}) = O(|\mathbf{z}|^{q-2p+1}) = O(|\mathbf{z}|^{2H_\tau+2-2p})$, $p = 0,1,2$, where $\Delta$ is the Laplace operator. These relations hold, because $\breve{f}_\varepsilon(\mathbf{z})$ is smooth function and

$$\varphi(\mathbf{z}) = |z|^{2H_\tau}[\breve{f}_\varepsilon(\mathbf{z}) - \breve{f}_\varepsilon(\mathbf{0}) - (\nabla\breve{f}(\mathbf{0}),\mathbf{z})]. \tag{A94}$$

Therefore

$$I_{\dot\Sigma}(\omega) = 2\pi L(2H_\tau)|\omega\mathbf{d}|^{-2-2H_\tau}\breve{f}_\varepsilon(\mathbf{0}) + o(|\omega\mathbf{d}|^{-4}). \tag{A95}$$

Formulas (A88), (A95) give the contribution of the inner points of $\Sigma$ in the HF asymptotics of the signal. The final formulas are (28-29).

***Contribution of*** $\partial\Sigma$.

Similarly to the way we got (A77), we find the contribution of the source boundary in (A88):

$$I_{\partial\Sigma}(\omega|\phi) = |\omega\mathbf{d}|^{-3}\sum_{\alpha,\beta} 2\pi m_\tau(\mathbf{x}_\alpha,\mathbf{x}_\beta)|\kappa_{\mathbf{x}_\alpha}\kappa_{\mathbf{x}_\beta}|^{-1/2}\varepsilon_{\mathbf{x}_\alpha}\bar{\varepsilon}_{\mathbf{x}_\beta} \ . \tag{A96}$$

Here $\mathbf{x}_\alpha = \mathbf{x}_\alpha(\mathbf{d})$ is an isolated boundary point such that $(\mathbf{x}-\mathbf{x}_\alpha,\mathbf{d}) = 0$ is a line tangent to $\partial\Sigma$ at $\mathbf{x}_\alpha$, $\kappa_{\mathbf{x}_\alpha}$ is the curvature of $\partial\Sigma$ at $\mathbf{x}_\alpha$. Unlike (A77), the summation in (A96) involves all pairs of the critical points, $(\mathbf{x}_\alpha,\mathbf{x}_\beta)$. This is correct for the smooth component of $m_\tau(\mathbf{x},\mathbf{y})$, i.e., for $\mu_\tau(\mathbf{x},\mathbf{y})$. The second component of $m_\tau(\mathbf{x},\mathbf{y})$ is smooth in $\Sigma\times\Sigma\setminus D$ only but it is continuous in $\Sigma\times\Sigma$ and vanishes on the diagonal $D$. Therefore the contribution of the second component and the point $(\mathbf{x}_\alpha,\mathbf{x}_\alpha)$ in (A96) is zero. As above (see section "The case $\mathbf{x}_0 = \mathbf{y}_0$", Appendix 2) it is possible to show that the contribution of the second component of $m_\tau(\mathbf{x},\mathbf{y})$ and any point $(\mathbf{x}_\alpha, \mathbf{x}_\alpha)$ in (A88) is considerably below (A95). Therefore (28-29) contains the complete information on the leader terms in the HP asymptotics of the signal.

## A4. Model B.

We assume that $\delta t_r(\mathbf{x}) = \xi(\mathbf{x})$ , and $\xi(\mathbf{x})$ is fractal in sense (32). If the random vector $(\xi(\mathbf{x}), \xi(\mathbf{y}))$ is Gaussian with zero mean, then $\xi(\mathbf{x}) - \xi(\mathbf{y})$ is a Gaussian variable with variance $\Delta_\xi^2(\mathbf{x}, \mathbf{y})$ and

$$\varphi_\Delta(\omega|\mathbf{x}, \mathbf{y}) = E\exp\{i\omega(\xi(\mathbf{x}) - \xi(\mathbf{y})\} = e^{-\omega^2\Delta_\xi^2(\mathbf{x},\mathbf{y})/2} \tag{A97}$$

Therefore, one has

$$A^{-2}\omega^2 E|\hat{u}(\omega)|^2 = \iint_{\Sigma\times\Sigma} m_\tau(\mathbf{x}, \mathbf{y})\exp(-i\omega((\mathbf{x}-\mathbf{y}), \mathbf{d}) - \omega^2\Delta_\xi^2(\mathbf{x}, \mathbf{y})/2)\sigma(d\mathbf{x})\sigma(d\mathbf{y}) \doteq I(\omega). \tag{A98}$$

By (32),

$$\Delta_\xi^2(\mathbf{x}, \mathbf{y}) \approx \alpha_\xi^2(\mathbf{x})|\mathbf{x}-\mathbf{y}|^{2H_\xi}, \qquad \mathbf{x}-\mathbf{y} \to \mathbf{0}, \tag{A99}$$

and $\alpha_\xi^2(\mathbf{x}) > c$ . Hence, $c < \Delta_\xi^2(\mathbf{x}, \mathbf{y})/|\mathbf{x}-\mathbf{y}|^{2H_\xi} < C$ . If $\Delta_\xi^2 > 0$, the term $\exp(-\omega^2\Delta_\xi^2/2)$ is rapidly decreasing function of $\omega$. Therefore, the main contribution into the asymptotics of $I(\omega), \omega >> 1$ comes from a small vicinity of the diagonal $\mathbf{x}-\mathbf{y} = \mathbf{0}$. But near the diagonal we have $m_\tau(\mathbf{x}, \mathbf{y}) \approx E\tau^2(\mathbf{x})$ and $\exp(i\omega(\mathbf{x}-\mathbf{y})) \approx 1$.Therefore

$$I(\omega) = \iint_{\Sigma\times\Sigma} E\tau^2(\mathbf{x})\exp(-\omega^2\alpha_\xi^2(\mathbf{x})|\mathbf{x}-\mathbf{y}|^{2H_\xi}/2)\chi(\varepsilon - |\mathbf{x}-\mathbf{y}|)\sigma(d\mathbf{x})\sigma(d\mathbf{y}) + \delta_\varepsilon , \tag{A100}$$

where $\chi(\cdot)$ is the Heaviside step function, $\varepsilon$ is small enough. If $\varepsilon = (c^{-1}\omega^{-1}\ln\omega)^{1/H_r}$, then

$\delta_\varepsilon = O(\omega^{-N})$ where $N > .5\ln\omega$.

Changing the variable: $(\mathbf{x}, \mathbf{y}) \to (\mathbf{x}, \mathbf{z} = \mathbf{x}-\mathbf{y})$ and using notation $r = |z|$, we can continue:

$$I(\omega) \approx 2\pi\int_\Sigma E\tau^2(\mathbf{x})\int_0^\varepsilon \exp(-0.5\omega^2\alpha_\xi^2(\mathbf{x})r^{2H_\xi})rdr\sigma(d\mathbf{x})$$

$$\approx 2\pi\int_\Sigma E\tau^2(\mathbf{x})\alpha_\xi^{-2/H_\xi}(\mathbf{x})\sigma(d\mathbf{x})\int_0^R \exp(-0.5r^{2H_\xi})rdr\cdot\omega^{-2/H_\xi}, \tag{A101}$$

where

$$R = \varepsilon(\alpha_\xi(\mathbf{x})\omega)^{1/H_r} = (c^{-1}\alpha_\xi(\mathbf{x})\ln\omega)^{1/H_r} > (\ln\omega)^{1/H_r} >> 1, \quad \omega >> 1. \tag{A102}$$

Setting $R = \infty$, one has

$$A^{-2}\omega^2 E|\hat{u}(\omega)|^2 \approx \pi 2^{1/H_\xi}\Gamma(1+1/H_\xi)\int_\Sigma E\tau^2(\mathbf{x})\alpha_\xi^{-2/H_\xi}(\mathbf{x})\sigma(d\mathbf{x})\cdot\omega^{-2/H_\xi}. \quad \text{(A103)}$$

This relation implies (33, 34).

Suppose that $\xi(\mathbf{x})$ has the non-isotropic fractal property (35) with indexes: $\mathbf{H}_r = (H', H'')$, i.e.

$$c_-^2 < \Delta_\xi^2(\mathbf{x},\mathbf{y})/(|x'-y'|^{2H'} + |x''-y''|^{2H''}) < c_+^2 . \quad \text{(A104)}$$

As above

$$I(\omega) \approx \iint_{\Sigma\times\Sigma} E\tau^2(\mathbf{x})\exp(-\omega^2\Delta_\xi^2(\mathbf{x},\mathbf{y})/2)\chi(\varepsilon - |\mathbf{x}-\mathbf{y}|)\sigma(d\mathbf{x})\sigma(d\mathbf{y}) \doteq \tilde{I}(\omega) \quad . \quad \text{(A105)}$$

Suppose that $\Sigma = (0,L)\times(0,W)$ and $\mathrm{M}_- < m_\tau(\mathbf{x},\mathbf{y}) < \mathrm{M}_+$. Let us estimate $\tilde{I}(\omega)$ from above. By (A105)), one has

$$\tilde{I}(\omega) < \mathrm{M}_+ \int_0^L\int_0^L \exp(-\omega^2 c_-^2|x'-y'|^{2H'})\chi(\varepsilon - |x_1 - y_1|)dx'dy'$$

$$\times\int_0^W\int_0^W \exp(-\omega^2 c_-^2|x''-y''|^{2H''}\chi(\varepsilon - |x''-y''|^{H''})dx''dy'' . \quad \text{(A106)}$$

Replacing $\mathrm{M}_+$ by $\mathrm{M}_-$ and $c_-$ by $c_+$, we get the similar estimate from below.

The resulting integral estimates look like (A101). Therefore, proceeding similarly, we arrive at the following relation: $\tilde{I}(\omega) = O(\omega^{-(H_1^{-1}+H_2^{-1})})$. It means that

$$E|\hat{u}(\omega)|^2 = O(\omega^{-2-2/\overline{H}_\xi}), \overline{H}_\xi = 2/(1/H' + 1/H''). \quad \text{(A107)}$$

## A5.Nucleation point: contribution into the spectral asymptotics.

The contribution of the nucleation point $\mathbf{x}_0$ in the displacement spectra is given by (49). We have to specify this relation depending on the stochastic nature of $\delta t_r(\mathbf{x})$ .

***Model A***: $\delta t_r(\mathbf{x}) = \xi^2(\mathbf{x}), (\xi(\mathbf{x}), \tau(\mathbf{x}))$ satisfies conditions (C1-C3).

Similarly to (A15), one

has $A^{-2}\omega^3 E|\hat{u}_{\mathbf{x}_0}(\omega)|^2 \approx \iint \phi(\mathbf{x},\mathbf{y})\exp(-i\omega((\mathbf{x}-\mathbf{y}),\boldsymbol{\gamma})c^{-1} + i\omega v^{-1}(|\mathbf{x}|-|\mathbf{y}|))\sigma(d\mathbf{x})\sigma(d\mathbf{y}) \doteq I(\omega)$.

(A108)

Here

$$\phi(\mathbf{x},\mathbf{y}) = \varphi(\mathbf{x})\varphi(\mathbf{y})\ m_\tau(\mathbf{x},\mathbf{y})/b_\xi(\mathbf{x},\mathbf{y}), \quad \text{(A109)}$$

where $m_\tau(\mathbf{x},\mathbf{y}), b_\xi(\mathbf{x},\mathbf{y})$ are given by (A13) and (A16) respectively, and $\varphi(\mathbf{x})$ is a smooth finite function such that $\varphi(\mathbf{x}) = 0$ outside of a small vicinity of $\mathbf{x}_0$ and $\varphi(\mathbf{x}_0) = 1$. In this case

$$m_\tau(\mathbf{x},\mathbf{y})/b_\xi(\mathbf{x},\mathbf{y}) \approx 0.5E\tau^2(\mathbf{x}_0)\sigma_\xi^{-1}(\mathbf{x}_0)\alpha_\xi^{-1}(\mathbf{x}_0)|\mathbf{x}-\mathbf{y}|^{-H_\xi} \doteq B^2(\mathbf{x}_0)|\mathbf{x}-\mathbf{y}|^{-H_\xi}. \qquad \text{(A110)}$$

Hence

$$\widetilde{A}^{-2}\omega^3 E|u_{\mathbf{x}_0}(\omega)|^2 \approx \iint_{\Sigma\times\Sigma}\varphi(\mathbf{x})\varphi(\mathbf{y})|\mathbf{x}-\mathbf{y}|^{-H_\xi}\psi(\omega\mathbf{x})\psi^*(\omega\mathbf{y})\sigma(d\mathbf{x})\sigma(d\mathbf{y}) := I(\omega) \qquad \text{(A111)}$$

where $\widetilde{A} = |A|B(x_0)$, (*) denotes the complex conjugation, and

$$\psi(\mathbf{x}) = \exp(-i(\mathbf{x},\boldsymbol{\gamma})c^{-1} + i|\mathbf{x}|v^{-1}). \qquad \text{(A112)}$$

***Reduction of*** $I(\omega)$**.** As well known

$$\iint f_1(\mathbf{x})f_2^*(\mathbf{y})|\mathbf{x}-\mathbf{y}|^{-H}\sigma(d\mathbf{x})\sigma(d\mathbf{y}) = C_H\iint \hat{f}_1(\boldsymbol{\kappa})\hat{f}_2^*(\mathbf{k})|\mathbf{k}|^{-2+H}\sigma(d\mathbf{k})$$

where $\hat{f}(\mathbf{k})$ is Fourier transform of $f(\mathbf{x})$ and $C_H = L(-H)/(2\pi)$ is given by (27) (Gel'fand and Shilov, 1964). If $\varphi(\mathbf{x}) = \widetilde{\varphi}(|\mathbf{x}-\mathbf{x}_0|)$ and $\mathbf{x}_0 = 0$, then

$$I(\omega) = C_{H_\xi}\int |J(\mathbf{k})|^2|\mathbf{k}|^{-2+H_\xi}\sigma(d\mathbf{k}), \qquad \text{(A113)}$$

where

$$J(\mathbf{k}) = \int e^{i(\mathbf{k},\mathbf{x})}\varphi(\mathbf{x})\psi(\mathbf{x}\omega)\sigma(d\mathbf{x}) =$$

$$\int_{-\pi}^{\pi}d\theta_x\int_0^\infty r\widetilde{\varphi}(r)\exp(ir(|k|\cos(\theta_x-\theta_k)+(v^{-1}-c^{-1}\cos\theta_x)\omega))dr. \qquad \text{(A114)}$$

Here $(r,\theta_{\mathbf{x}}),(|\mathbf{k}|,\theta_{\mathbf{k}})$ are polar coordinates of $\mathbf{x}$ and $\mathbf{k}$, respectively; the coordinate system $(r,\theta_{\mathbf{x}})$ is such that $\boldsymbol{\gamma} = (1,0)$.

Setting

$$F(u) = \int_0^\infty e^{iur}r\widetilde{\varphi}(r)dr \quad \text{and} \quad |\mathbf{k}| = \rho\,\omega, \qquad \text{(A115)}$$

one has

$$I(\omega) = C_{H_\xi}\omega^{H_\xi}\widetilde{I}(\omega), \qquad \text{(A116)}$$

where

$$\widetilde{I}(\omega) = \int_0^\infty \rho^{-1+H_\xi}d\rho\int_{-\pi}^{\pi}|J(\rho,\theta_k)|^2 d\theta_k \qquad \text{(A117)}$$

$$J(\rho,\theta_k)=\int_{-\pi}^{\pi}F(\omega H(\rho,\theta_k,\theta_x)d\theta_x \qquad \text{(A118)}$$

$$H(\rho,\theta_k,\theta_x)=\cos(\theta_x-\theta_k)\rho+(v^{-1}-c^{-1}\cos\theta_x) \ . \qquad \text{(A119)}$$

By change of variables: $(\rho,\theta_k)\Rightarrow(\tilde{\rho},\theta)$, such that $\rho e^{-i\theta_k}-c^{-1}=-v^{-1}\tilde{\rho}e^{-i\theta}$, one has

$$H(\rho,\theta_k,\theta_x)=v^{-1}(1-\tilde{\rho}\cos(\theta_x-\theta)),$$

$$\tilde{I}(\omega)=2v^{-2}\int_0^{\infty}\int_0^{\pi}\left|c^{-1}-v^{-1}\tilde{\rho}e^{-i\theta}\right|^{-2+H_\xi}\tilde{\rho}\left|\tilde{J}(\tilde{\rho}|\omega)\right|^2 d\tilde{\rho}d\theta \ , \qquad \text{(A120)}$$

$$\tilde{J}(\tilde{\rho}|\omega)=\int_{-\pi}^{\pi}F(\omega v^{-1}(1-\tilde{\rho}\cos(\theta_x-\theta))d\theta_x=2\int_0^{\pi}F(\omega v^{-1}(1-\tilde{\rho}\cos\alpha))d\alpha \qquad \text{(A121)}$$

To analyze $\tilde{I}(\omega)$ for large $\omega$, note that

$$F(u)=\begin{cases}-u^{-2}(1+o(1)),u\to\infty\\ F(0)+O(|u|),u\to 0\end{cases} ; \qquad \text{(A122)}$$

in addition,

$$f(\tilde{\rho},\theta)\doteq\tilde{\rho}\left|c^{-1}-v^{-1}\tilde{\rho}e^{-i\theta}\right|^{-2+H_\xi} \qquad \text{(A123)}$$

is locally (but not globally) integrable because $f(\tilde{\rho},\theta)d\tilde{\rho}d\theta=\rho^{-1+H_\xi}d\rho d\theta_k$

and $f(\tilde{\rho},\theta)=O(\tilde{\rho}^{-1+H_\xi}),\tilde{\rho}\to\infty$. These properties and (A121) result in fact that the main contribution in the asymptotics of $\tilde{I}(\omega)$ comes from a small vicinity of the curve: $\cos\alpha-1/\tilde{\rho}=0$, i.e. from the set

$$\Omega_\varepsilon=\{(\tilde{\rho},\alpha):\left|\cos\alpha-\tilde{\rho}^{-1}\right|<\varepsilon,0\le\alpha\le\pi/2\}.$$

This can be seen as follows. Suppose $(\tilde{\rho},\alpha)$ belongs to the supplement of $\Omega_\varepsilon$, i.e. to $\Omega_\varepsilon^c$. Then for large $\tilde{\rho}\omega$ one has $F(\omega v^{-1}(1-\tilde{\rho}\cos\alpha))=O(\tilde{\rho}^{-2}\omega^{-2})$. Since $f(\tilde{\rho},\theta)\tilde{\rho}^{-2}$ is integrable in vicinity of $\tilde{\rho}=\infty$, we conclude from (A120, A121) that the contribution of $\Omega_\varepsilon^c$ in the asymptotics of $\tilde{I}(\omega)$ is

$$\tilde{I}_{\Omega_\varepsilon^c}(\omega)=O(\omega^{-4}). \qquad \text{(A124)}$$

***The contribution of*** $\Omega'_\varepsilon=\Omega_\varepsilon\cap\{\tilde{\rho}>1+\varepsilon\}$.

If $0\le\alpha\le\pi/2$ and $\tilde{\rho}\ge 1$, the equation $\cos\alpha-1/\tilde{\rho}=0$ has unique solution $\alpha_0=\arccos(1/\tilde{\rho})$.

For small $\varepsilon$ and $(\tilde{\rho},\alpha)\in\Omega'_\varepsilon$

$$1-\tilde{\rho}\cos\alpha=\tilde{\rho}(\cos\alpha_0-\cos\alpha)\approx\sqrt{\tilde{\rho}^2-1}(\alpha-\alpha_0).$$

The contribution of $\Omega'_\varepsilon$ in the asymptotics of $\widetilde{I}(\omega)$ is

$$\widetilde{I}_{\Omega'_\varepsilon}(\omega)=2v^{-2}\int_{1+\varepsilon}^{\infty}\int_0^{\pi}\left|c^{-1}-v^{-1}\widetilde{\rho}e^{-i\theta}\right|^{-2+H_\xi}\widetilde{\rho}\left|\widetilde{J}_\varepsilon(\widetilde{\rho}|\omega)\right|^2 d\widetilde{\rho}d\theta \tag{A125}$$

where

$$\widetilde{J}_\varepsilon(\widetilde{\rho}|\omega)=2\int_{\alpha_0-\varepsilon}^{\alpha_0+\varepsilon}F(v^{-1}\sqrt{\widetilde{\rho}^2-1}(\alpha-\alpha_0)\omega)d\alpha=2v(\omega\sqrt{\widetilde{\rho}^2-1})^{-1}\int_{-a\omega}^{a\omega}F(u)du \tag{A126}$$

and $$a=\varepsilon v^{-1}\sqrt{\widetilde{\rho}^2-1}>\varepsilon^{3/2}v^{-1}\sqrt{2}\,.$$

By (A115),

$$\int_{-a\omega}^{a\omega}F(u)du=-i\int_0^{\infty}(e^{ia\omega r}-e^{-ia\omega r})\widetilde{\varphi}(r)dr=2\widetilde{\varphi}(0)(a\omega)^{-1}(1+o(1)),\omega\to\infty\,, \tag{A127}$$

where $\widetilde{\varphi}(0)=1$. Substituting (A126, A127) in (A125), we obtain

$$\widetilde{I}_{\Omega'_\varepsilon}(\omega)=O(\omega^{-4}) \tag{A128}$$

because $f(\widetilde{\rho},\theta)/(\widetilde{\rho}^2-1)^2$ is integrable on $\{\widetilde{\rho}\ge 1+\varepsilon,\theta\in(0,\pi)\}$.

***The contribution of*** $\Omega^0_\varepsilon=\Omega_\varepsilon\cap\{|\widetilde{\rho}-1|<\varepsilon\}$.

Set $\Omega^0_\varepsilon$ is a small vicinity of $(\widetilde{\rho},\alpha)=(1,0)$ .Therefore

$$1-\widetilde{\rho}\cos\alpha\approx 1-\widetilde{\rho}+\alpha^2/2,\ (\widetilde{\rho},\alpha)\in\Omega^0_\varepsilon$$

and the contribution of $\Omega^0_\varepsilon$ in the asymptotics of $\widetilde{I}(\omega)$ is

$$\widetilde{I}_{\Omega^0_\varepsilon}(\omega)=8v^{-2}\int_0^{\pi}\left|c^{-1}-v^{-1}e^{i\theta}\right|^{-2+H_\xi}d\theta U_\varepsilon(\omega), \tag{A129}$$

$$U_\varepsilon(\omega)=\int_{-1+\varepsilon}^{1+\varepsilon}d\widetilde{\rho}\left|\int_{-\varepsilon}^{\varepsilon}d\alpha F(v^{-1}(\omega(1-\widetilde{\rho})+(\sqrt{\omega}\alpha)^2/2)\right|^2.$$

By change of variables: $v^{-1}\omega(1-\widetilde{\rho})=x,\sqrt{\omega}\alpha=\widetilde{\alpha}$ , we obtain

$$U_\varepsilon(\omega)=v\omega^{-2}\int_{-\varepsilon\omega}^{\varepsilon\omega}dx\left|\int_{-\varepsilon\omega^{1/2}}^{\varepsilon\omega^{1/2}}d\widetilde{\alpha}F(x+\widetilde{\alpha}^2/2)\right|^2\doteq V_\varepsilon(\omega)\omega^{-2} \tag{A130}$$

To find the limit value of $V_\varepsilon(\omega)$ as $\omega\to\infty$ , note that

$$\int_{-A}^{A}d\widetilde{\alpha}F(x+\widetilde{\alpha}^2/2)=\int_0^{\infty}dre^{ixr}r\widetilde{\varphi}(r)\int_{-A}^{A}e^{ir\widetilde{\alpha}^2/2}d\widetilde{\alpha},\quad A=\varepsilon\sqrt{\omega}\,.$$

Using integration by parts we can continue as follows:

$$=2(iA)^{-1}\int_0^{\infty}(e^{iA^2r/2}-1)\widetilde{\varphi}(r)dr+\int_0^{\infty}[\int_{-A}^{A}d\widetilde{\alpha}(e^{ir\widetilde{\alpha}^2/2}-1)/(i\widetilde{\alpha}^2)]e^{ixr}\widetilde{\varphi}(r)dr$$

$$= O(A^{-1}) + \int_0^{\infty} e^{ixr} \tilde{\varphi}(r) \Phi_A(r) dr$$

where

$$\Phi_A(r) = \int_{-A}^{A} d\alpha (e^{ir\alpha^2/2} - 1)/(i\alpha^2) = -i\sqrt{2r} \int_0^{A^2 r/2} (e^{iv} - 1) v^{-3/2} dv = \sqrt{2\pi r} e^{i\pi/4} + qA^{-1}, \quad |q| \le 8 .$$

Hence,

$$\lim_{\omega \to \infty} V_\varepsilon(\omega) = 2\pi v \int_{-\infty}^{\infty} dx \left| \int_0^{\infty} e^{ixr} \tilde{\varphi}(r) \sqrt{r} dr \right|^2 = (2\pi)^2 v \int_0^{\infty} \tilde{\varphi}^2(r) r dr = 2\pi v \int \varphi^2(\mathbf{x}) \sigma(d\mathbf{x}) \quad \text{(A131)}$$

By (A129-A131),

$$\tilde{I}_{\Omega_\varepsilon^0}(\omega) \approx 2\pi C_\gamma \omega^{-2} \int \varphi^2(\mathbf{x}) \sigma(d\mathbf{x}), \quad \text{(A132)}$$

$$C_\gamma = 4 v^{-1} \int_{-\pi}^{\pi} \left| c^{-1} - v^{-1} e^{i\theta} \right|^{-2+H_\xi} d\theta \quad \text{(A133)}$$

Comparing (A124), (A128), (A132) and using (A111), (A116) we conclude that

$$r.m.s.u_{\mathbf{x}_0}(\omega) \approx |A| B(\mathbf{x}_0) \left| L(-H_\xi) C_\gamma \right|^{1/2} \|\varphi\| \omega^{-2-(1-H_\xi)/2}, \quad \text{(A134)}$$

where $B^2(\mathbf{x}_0) = 0.5 E\tau^2(\mathbf{x}_0) \sigma_\xi^{-1}(\mathbf{x}_0) \alpha_\xi^{-1}(\mathbf{x}_0)$ and $L(h)$ is given by (27).

***Model LF***: $\delta t_r(\mathbf{x}) = 0$, $\tau(\mathbf{x})$ satisfies conditions (C2, C3), $\mathbf{x}_0 = 0$.

By (49),

$$A^{-2} \omega^2 E \left| \hat{u}_{\mathbf{x}_0}(\omega) \right|^2 =$$

$$\iint \phi(\mathbf{x}, \mathbf{y}) \exp(-i\omega((\mathbf{x} - \mathbf{y}), \boldsymbol{\gamma}) c^{-1} + i\omega v^{-1}(|\mathbf{x}| - |\mathbf{y}|)) \sigma(d\mathbf{x}) \sigma(d\mathbf{y}) \doteq I(\omega|\phi), \quad \text{(A135)}$$

where $\phi = \phi_1 + \phi_2$,

$$\phi_1 = \varphi(\mathbf{x}) m_\tau(\mathbf{x}) \varphi(\mathbf{y}) m_\tau(\mathbf{y}), \qquad m_\tau(\mathbf{x}) = E\tau(\mathbf{x}), \quad \text{(A136)}$$

$$\phi_2 \approx \alpha_\tau^2(\mathbf{x}_0) |\mathbf{x} - \mathbf{y}|^{2H_\tau}, \quad \text{(A137)}$$

$\varphi(\mathbf{x}) = \tilde{\varphi}(|\mathbf{x}|)$ is the same as in (A109).

***Asimptotics of*** $I(\omega|\phi_1)$. By (A136), $I(\omega|\phi_1) = |J(\omega|a)|^2$, where $a = \varphi(\mathbf{x}) m_\tau(\mathbf{x})$ is a smooth finite function and

$$J(\omega|a) = \int a(\mathbf{x}) e^{i\omega v^{-1}|\mathbf{x}| h(\varphi)} \sigma(d\mathbf{x}),$$

$$h(\varphi) = v^{-1} - (\mathbf{x}/|\mathbf{x}|, \boldsymbol{\gamma}) c^{-1} = v^{-1} - c^{-1} \cos\varphi .$$

Note, that $h(\varphi) \ne 0$ because $v < c$. Using the polar coordinates: $\mathbf{x} = re^{i\varphi}$, one has

$$J(\omega|a) = \int_{-\pi}^{\pi} d\varphi \int_0^{\infty} ra(re^{i\varphi})e^{i\omega v^{-1}rh(\varphi)}dr \ . \tag{A138}$$

Double integration over $r$ by parts leads to the following relation

$$J(\omega|a) = a(0)\omega^{-2}\int_{-\pi}^{\pi}(v^{-1} - c^{-1}\cos\varphi)^{-2}d\varphi + o(\omega^{-2})\,. \tag{A139}$$

Hence,

$$I(\omega|\phi_1) = O(\omega^{-4})\,. \tag{A140}$$

***Asimptotics of*** $I(\omega|\phi_2)$. One has

$$I(\omega|\phi_2) = \alpha_\tau^2(\mathbf{x}_0)\iint b(\mathbf{x})b^{\bullet}(\mathbf{y})|\mathbf{x}-\mathbf{y}|^{2H_\tau}\sigma(d\mathbf{x})\sigma(d\mathbf{y})\,,$$

where $b(\mathbf{x}) = \varphi(\mathbf{x})\psi(\omega\mathbf{x})$ and $\psi$ is given by (A12).

If $0 < H < 1$, then

$$|\mathbf{x}|^{2H} + |\mathbf{y}|^{2H} - |\mathbf{x}-\mathbf{y}|^{2H} = L(2H)(2\pi)^{-1}\int(e^{i(\mathbf{k},\mathbf{x})}-1)(e^{-i(\mathbf{k},\mathbf{y})}-1)|\mathbf{k}|^{-2-2H}\sigma(d\mathbf{k})$$

(Gel'fand and Shilov, 1964). Therefore

$$I(\omega|\phi_2) = B_H(\mathbf{x}_0)\int|J(\mathbf{k}) - J(\mathbf{0})|^2|\mathbf{k}|^{-2-2H_\tau}\sigma(d\mathbf{k}) + R(\omega)\,, \tag{A141}$$

where $B_H(\mathbf{x}_0) = -\alpha_\tau^2(\mathbf{x}_0)L(2H_\tau)/(2\pi)$ and $J(\mathbf{k}) = \hat{b}(\mathbf{k})$ is given by (A114);

$$R(\omega) = -2\operatorname{Re}J(\omega|b)J^{\bullet}(\omega|b_H)\,, \qquad b_H = |x|^{2H}b(x)\,,$$

$J(\omega|a)$ is given by (A138).

By (A139), $J(\omega|b) = O(\omega^{-2})$. Similarly to (A138),

$$J(\omega|b_H) = \int_{-\pi}^{\pi} d\varphi\int_0^{\infty} r^{1+2H_\tau}b(re^{i\varphi})e^{i\omega v^{-1}rh(\varphi)}dr$$

Using double integration over $r$ by parts and the result (A5), we obtain

$$J(\omega|b_H) = -b(0)\Gamma(2+2H_\tau)e^{i\pi H_\tau}\int_{-\pi}^{\pi}(v^{-1} - c^{-1}\cos\varphi)^{-2-2H_\tau}d\varphi\omega^{-2-2H_\tau}(!+o(1))\,.$$

Hence,

$$R(\omega) = -2\operatorname{Re}J(\omega|b)J^{\bullet}(\omega|b_H) = O(\omega^{-4-2H_\tau}). \tag{A142}$$

Let us consider now the main component of $I(\omega|\phi_2)$, namely

$$I_1(\omega) = \int|J(\mathbf{k}) - J(\mathbf{0})|^2|\mathbf{k}|^{-2-2H_\tau}\sigma(d\mathbf{k})\,. \tag{A143}$$

This term looks like (A113). Therefore, in a similar manner to (A116), (A120), we have

$$I_1(\omega) = \omega^{-2H_\tau} \widetilde{I}_1(\omega), \tag{A144}$$

$$\widetilde{I}_1(\omega) = 2v^{-2} \int_0^\infty \int_0^\pi \left| c^{-1} - v^{-1} \widetilde{\rho} e^{-i\theta} \right|^{-2-2H_\tau} \widetilde{\rho} \left| \widetilde{J}(\widetilde{\rho}|\omega) - \widetilde{J}(\widetilde{\rho}_0|\omega) \right|^2 d\widetilde{\rho} d\theta, \tag{A145}$$

where $\widetilde{\rho}_0 = v / c$,

$$\widetilde{J}(\widetilde{\rho}|\omega) = 2\int_0^\pi F(\omega v^{-1}(1 - \widetilde{\rho}\cos\alpha)) d\alpha, \tag{A146}$$

and $F(\cdot)$ is given by (A115), (A122).

By (A122),

$$\widetilde{J}(\widetilde{\rho}_0|\omega) = -2\int_0^\pi v^2 (1 - \widetilde{\rho}_0 \cos\alpha)^{-2} d\alpha \cdot \omega^{-2}(1 + o(1)). \tag{A147}$$

Now consider

$$\widetilde{I}_{\Omega_\varepsilon}(\omega) = 2v^{-2} \iint_{\Omega_\varepsilon} \left| c^{-1} - v^{-1} e^{-i\theta} \right|^{-2-2H\tau} \widetilde{\rho} \left| \widetilde{J}(\widetilde{\rho}|\omega) \right|^2 d\widetilde{\rho} d\theta$$

where $\Omega_\varepsilon = \{ |\rho - \rho_0| > \varepsilon, 0 \le \theta \le \pi \}$. Repeating all previous steps (A124-A134), we conclude that

$$\widetilde{I}_{\Omega_\varepsilon}(\omega) = O(\omega^{-2}) \|\varphi\|^2. \tag{A148}$$

Combining (A143, A147, A148), we conclude that the contribution of the area $\Omega_\varepsilon$ in the asymptotics of $\widetilde{I}_1(\omega)$ is given by (A148).

Thus, it suffices to consider

$$\widetilde{I}_{\Omega_\varepsilon^c}(\omega) = 2v^{-2} \iint_{\Omega_\varepsilon^c} \left| c^{-1} - v^{-1} e^{-i\theta} \right|^{-2-2H\tau} \widetilde{\rho} \left| \widetilde{J}(\widetilde{\rho}|\omega) - \widetilde{J}(\rho_0|\omega) \right|^2 d\widetilde{\rho} d\theta$$

where $\Omega_\varepsilon^c = \{ |\rho - \rho_0| < \varepsilon, 0 \le \theta \le \pi \}$.

By (A122), (A146), in a vicinity of $\widetilde{\rho}_0$

$$\widetilde{J}(\widetilde{\rho}|\omega) - \widetilde{J}(\rho_0|\omega) \approx 2\int_0^\pi [(\omega v^{-1}(1 - \widetilde{\rho}\cos\alpha))^{-2} - (\omega v^{-1}(1 - \widetilde{\rho}_0 \cos\alpha))^{-2}) d\alpha$$

$$= 2v^2 \omega^{-2} \int_0^\pi \cos\alpha (1 - \widetilde{\rho}_0 \cos\alpha))^{-3}) d\alpha (\widetilde{\rho} - \widetilde{\rho}_0)$$

Therefore,

$$\widetilde{I}_{\Omega_\varepsilon^c}(\omega) = O(\omega^{-4}) \tag{A149}$$

because a singularity of the function $\left|c^{-1}-v^{-1}\cos\theta\right|^{-2-2H\tau}\left|\widetilde{\rho}-v/c\right|^{2}$ at the point $(\widetilde{\rho}.\theta)=(v/c,0)$ is integrable.

Combining (A144, A128, A149), we obtain

$$I_1(\omega)=\omega^{-2H_\tau}\widetilde{I}_1(\omega)=O(\omega^{-2-2H})\|\varphi\|^2$$

Taking into account (A135)) and (A142), we get the final result: $r.m.s.u_{\mathbf{x}_0}(\omega)\approx\|\varphi\|O(\omega^{-2-H_\tau})$.

***Model B***: $\delta t_r(\mathbf{x})=\xi(\mathbf{x})$, $(\xi(\mathbf{x}),\tau(\mathbf{x}))$ is the same as in model **B**. In this case

$$A^{-2}\omega^2 E\left|\hat{u}(\omega)\right|^2$$

$$=\iint_{\Sigma\times\Sigma}\phi(\mathbf{x},\mathbf{y})\exp(-i\omega((\mathbf{x}-\mathbf{y}),\boldsymbol{\gamma})c^{-1}+i\omega(|x|-|y|)-\omega^2\Delta_\xi^2(\mathbf{x},\mathbf{y})/2)\sigma(d\mathbf{x})\sigma(d\mathbf{y})\doteq I(\omega) \quad \text{(A150)}$$

where $\phi(\mathbf{x},\mathbf{y})=\varphi(\mathbf{x})\varphi(\mathbf{y})\, m_\tau(\mathbf{x},\mathbf{y})$, and $\Delta_\xi^2(\mathbf{x},\mathbf{y})\approx\alpha_\xi^2(\mathbf{x})\left|\mathbf{x}-\mathbf{y}\right|^{2H_\xi}$, $\quad \mathbf{x}-\mathbf{y}\to\mathbf{0}$.

The integral (A150) is similar to (A98). The additional term $\omega(|\mathbf{x}|-|\mathbf{y}|)$ is much less $\omega^2\Delta_\xi^2(\mathbf{x},\mathbf{y})$, $\mathbf{x}\neq\mathbf{y}$. Therefore the asymptotic analysis (A150) and (A98) is the same.

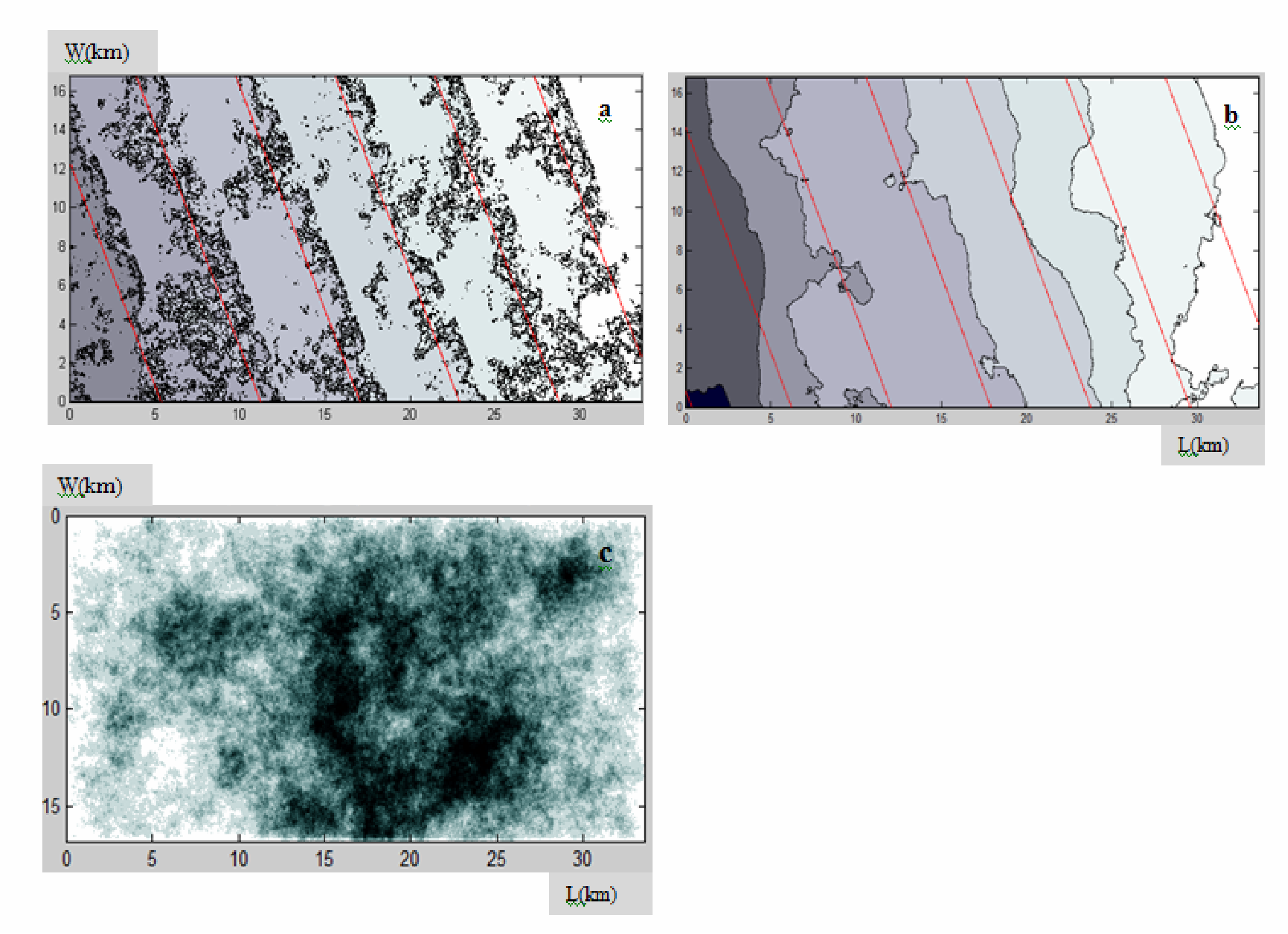

**Fig.1** The components ( $t_a(\mathbf{x}), \tau(\mathbf{x})$ ) of the stochastic earthquake source model (realizations).

(**a**) Arrival time function $t_a(\mathbf{x})$ in the form of isochrones for the model **A** with the smoothness indexes $H_r = 0.2$ ,

(**b)** the same as in (**a**) for $H_r = 0.8$ .The straight line isochrones correspond to the arrival time function $Et_a(\mathbf{x})$ ;

(**c**) the local stress drop function $\tau(\mathbf{x})$ with the smoothness index $H_\tau = 0.2$ .

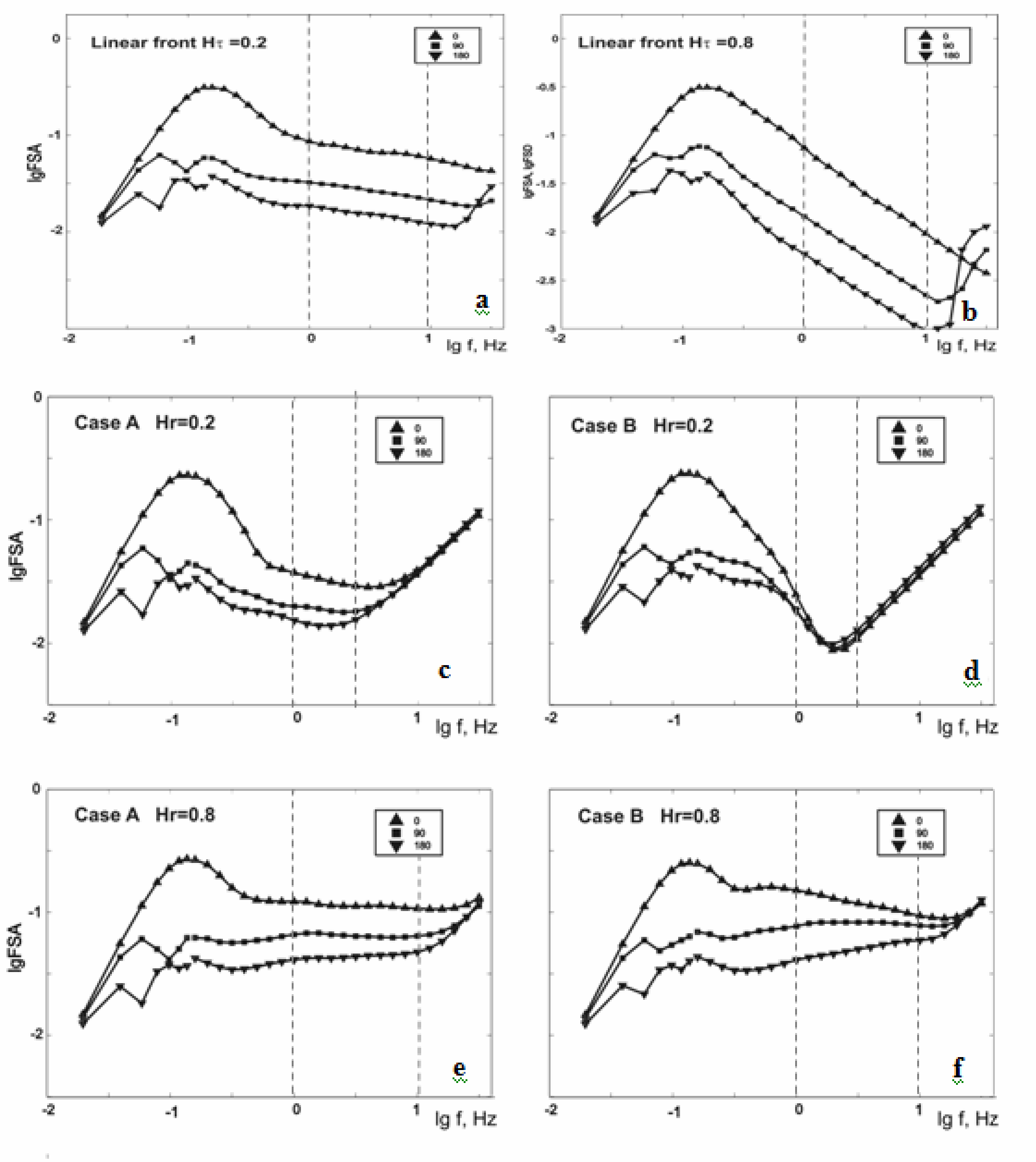

**Fig.2** R.M.S. spectra, $\hat{\hat{u}}(\omega)$ , for the doubly stochastic source models.

Each model is specified by vector ( *name,* $H_r$ , $H_\tau$ ) and by angle $\alpha$ , where *name* means **A**, **B,** or **LF(**linear front) model; $H_r$ is the fractal index of $t_a(\mathbf{x})$; $H_\tau$ is the fractal index of $\tau(\mathbf{x})$ ; $\alpha$ is angle between rays $\boldsymbol{\gamma}$ and $\boldsymbol{\gamma}_r$ : $\alpha = 0^o$ , $90^o$ , and. $180^o$

*Models*:(**a**) (**LF**,1,0.2), (**b**) (**LF**,1,0.8), (**c**) (**A**,0.2, 0.2), (**d**) (**B**,0,2,0,2), (**e**) (**A**,0.8,0.2) , (**f**) (**B,**0.8,0.2).

The dotted lines display the correct estimate of the spectra in the high (for the seismologist) frequency range: 1-10Hz (a, b, e, f), 1-3Hz (c, d).